\documentclass[12pt]{aastex}
\usepackage{emulateapj5}
\submitted{ApJ}
\usepackage{psfig}

\slugcomment{\apj , submitted}    %\slugcomment{Accepted by \apj}
\shorttitle{3D Identification and Reconstruction of Galaxy Groups}
\shortauthors{Marinoni {\it et al.~}}

\def\lsim{\raise0.3ex\hbox{$<$}\kern-0.75em{\lower0.65ex\hbox{$\sim$}}} 
\def\gsim{\raise0.3ex\hbox{$>$}\kern-0.75em{\lower0.65ex\hbox{$\sim$}}} 
\def\lesssim{\mathrel{\hbox{\rlap{\hbox{\lower4pt\hbox{$\sim$}}}\hbox{$<$}}}}
\def\gtrsim{\mathrel{\hbox{\rlap{\hbox{\lower4pt\hbox{$\sim$}}}\hbox{$>$}}}}
\def\sc1{\raise2.1ex\hbox{\tiny $r\!\!=\!\!4$}\kern-0.95em{\hbox{$=$}}}

\newcommand{\mamo}[1]{\mbox{$#1$}}
\newcommand{\unit}[1]{\ifmmode \:\mbox{\rm #1}\else \mbox{#1}\fi}
\newcommand{\sbr}[1]{_{\rm #1}}

\newcommand{\mone}{\mamo{^{-1}}}            %^{-1}

\newcommand{\kms}{\unit{km~s\mone}}

\newcommand{\mpc}{\unit{Mpc}}
\newcommand{\hmpc}{\mamo{h{\mone}\mpc }}     %h^-1 Mpc
\newcommand{\hinvmpcub}{\mamo{h^{-3} \mpc^{3}}}    %h^-3 Mpc^3
\def\ltsima{$\; \buildrel < \over \sim \;$}
\def\simlt{\lower.5ex\hbox{\ltsima}}
\def\gtsima{$\; \buildrel > \over \sim \;$}
\def\simgt{\lower.5ex\hbox{\gtsima}}          

\def\sc{{\rm Science\ }}

\begin{document}
\title{Three-Dimensional Identification and Reconstruction of  Galaxy Systems within
Deep Redshift Surveys}

\author{Christian Marinoni, Marc Davis,  Jeffrey A. Newman, and Alison
L. Coil} \affil{Department of Astronomy, University of California
at    Berkeley,     Berkeley,    CA    94720-3411,     US    \\e-mail:
{marinoni@astro.berkeley.edu,                  marc@astro.berkeley.edu,
jnewman@astro.berkeley.edu, acoil@astro.berkeley.edu}}

\begin{abstract}

We have developed a new geometrical method for identifying and
reconstructing a homogeneous and highly complete set of galaxy groups
in the next generation of deep, flux-limited redshift surveys.  Our
method combines information from the three-dimensional Voronoi diagram
and its dual, the Delaunay triangulation, to obtain group and cluster
catalogs that are remarkably robust over wide ranges in redshift and
degree of density enhancement.  As free byproducts, this
Voronoi-Delaunay method (VDM) provides a non-parametric measurement of
the galaxy density around each object observed and a quantitative
measure of the distribution of cosmological voids in the survey
volume.  In this paper, we describe the VDM algorithm in detail and test
its effectiveness using a family of mock catalogs that simulate the
DEEP2 Redshift Survey. We show that this survey will be quite suitable
for optically selecting distant clusters at $z\sim 1$ over a wide
range of richness.

Using the mock DEEP2 catalogs, we demonstrate that the VDM algorithm
can be used to identify a homogeneous set of groups in a
magnitude-limited sample ($I\sbr{AB}\le23.5$) throughout the survey
redshift window $0.7 < z < 1.2$.  The actual group membership can be
effectively reconstructed even in the distorted redshift space
environment for systems with line of sight velocity dispersion
$\sigma_{los}$ greater than $\approx 200$ \kms.  By applying the
sampling rate and the instrument-imposed target selection biases
expected for DEEP2, we show that we can construct a homogeneous sample
of systems which reproduces major properties of the ``real'' cluster
parent population down to $\approx 200$ \kms~ for systems with at
least 5 members.  In a $\Lambda$CDM cosmology this translates into an
identification rate of $\sim 270$ systems per square degree and a
total of more than 1000 groups within the full DEEP2 survey volume.

By comparing the galaxy cluster catalog derived from the mock DEEP2
observations to the underlying distribution of clusters found in real
space with much fainter galaxies included (which should more closely
trace mass in the cluster), we can assess completeness in velocity
dispersion directly.  We conclude that the recovered DEEP2 group and
cluster sample should be statistically complete for $\sigma_{los}
\gtrsim 400$ \kms.  Finally, we argue that the reconstructed bivariate
distribution of systems as a function of redshift and velocity
dispersion reproduces with high fidelity the underlying real space
distribution and can thus be used robustly to constrain cosmological
parameters.

\end{abstract} 

\keywords{ galaxies: high-redshift -- 
galaxies: clusters: general -- cosmology: large--scale structure of the
universe -- methods: data analysis}

% % Opzioni di stampa %
%\onecolumn  % Versione righe spaziate   
%\twocolumn  % Versione compressa

\section{Introduction} 

Non-linear galaxy overdensities provide useful cosmological probes,
particularly as objects ranging from small groups to rich clusters can
be described simultaneously using relatively simple empirical and
theoretical distributions. \citet{mh01}, for example, were able to
investigate the mapping between mass, light, and other cluster
observables for objects ranging from $\sim 10^{12}$ to $\sim 10^{15}
M_{\odot}$ in mass.  The abundance of groups and clusters itself is a
fundamental observable expected to have evolved substantially since
high redshift, with the strength of that evolution sensitively
dependent upon fundamental cosmological parameters \citep{lilje92,
bah97}. \citet{new01} have shown that this variant of the classical
dN/dz test can provide significant constraints on ``dark energy''
models when applied to upcoming redshift surveys.  However, a key
issue in whether we can use clusters to make precision cosmological
measurements is the availability of efficient and objective
cluster-finding algorithms.

In recent years there has been much work on detecting two-dimensional
galaxy overdensities in wide-field optical imaging surveys for
subsequent spectroscopic follow up \citep{sco99, gal00, gy00, gon01}.
However, these systematic searches for clusters, fueled by the
availability of large CCD camera mosaics, are in many cases biased
towards special subsets of the general galaxy population such as red
objects or AGNs. A closely related problem is that the recovered
cluster samples are statistically complete only near the upper tail of
the velocity dispersion distribution.  In effect, these methods
identify only those rich aggregates that are most conspicuous.
However, less extreme systems such as galaxy groups, which contain
most of the luminosity, and presumably mass, of the universe, may be
more useful probes of the large-scale structure.

With the next generation of multi-object spectrographs on large
telescopes soon to be installed, deep redshift surveys (e.g., the
DEEP2/DEIMOS Survey \citep{dav00}, hereafter referred to as DEEP2, and
the VLT/VIRMOS redshift survey \citep{lf01}) are well within reach.
It is thus worthwhile to address the more ambitious task of
constructing large, statistically complete samples of galaxy systems
selected over wide redshift baselines and covering a broad range in
density enhancement and velocity dispersion using the
three-dimensional redshift space data such surveys will provide.

Because  of  the high  spectroscopic  resolution  (FWHM  $\sim 65$  km
s$^{-1}$) and relatively dense sampling to be used, the DEEP2 Redshift
Survey  will   be  uniquely   suitable  for  measuring   the  velocity
dispersions and the internal dynamics of large numbers of distant 
galaxy clusters, and  selecting a group sample without relying on
non-kinematic properties  (X-ray brightness,
galaxy richness in optical images, etc.).
DEEP2   will  obtain  data characterizing  galaxies  and
large-scale  structure that are  comparable to  those provided  by the
best previous surveys of the  local universe, but for objects at high
redshift,  $z\sim1$.   Furthermore,  sensitive  S-Z  observations  are
planned for all DEEP2 fields, which will allow the virialization state
of clusters to be assessed.

If we are to use DEEP2 clusters to make cosmological measurements, we
must be able to reliably identify high redshift groups and clusters
and determine their members and properties robustly. However, rich
clusters are rare events; only $\sim 10$ Coma-like clusters are
expected in the DEEP2 survey volume. The mass function of clusters is
very steep, so to overcome Poisson noise detecting the most abundant
groups of lower mass is critical. 
To improve the predictive power of the cosmological tests, we must be
able to identify systems down to small, group-scale velocity
dispersions ($\sigma_{los}$) in a complete and unbiased way.

A variety of methods for identifying galaxy overdensities have been
applied to redshift surveys of the local, $z\sim0$ universe.  The
hierarchical \citep{mat78,tul80} and percolation (also known as
``friends-of-friends'', hereafter simply FOF) methods \citep{huc82} of
group identification have both been widely used (e.g.
\citep{tul87,mdl89,rgh89,hg91,gcf92,nol93,gar93,rpg97,tra98,ram99,giu00,tuc00}).  
Their main characteristics and
drawbacks are discussed in \citet{mariphd}.

These standard cluster-finding  algorithms are unsatisfactory for many
reasons.  For instance, the fixed-radius search window at the heart of
standard FOF techniques is insensitive to local variations
in  the  density of  points.   Assigned  cluster membership  therefore
depends  on  the scale  of  the adopted  linking  length  and not  the
distribution of galaxies alone, violating the dictum to ``let the data
speak for themselves'' \citep{ope84}.   In fact, both the hierarchical
and the percolation methods  require prior knowledge and/or user-fixed
parameters to produce their best results.  Density thresholds, linking
length parameter  scaling laws, galaxy selection  functions, etc. must
all  be set in advance.  The  pre-processing and/or  trial-and-error tests
required  to  tune  these  algorithms  for  a  particular  dataset  is
extremely  inefficient and  may  even lead  to systematic  differences
amongst different applications of the same technique.

It is  well known  that the performance  of the  standard FOF
algorithm across a  wide range of density enhancements  is not uniform
(Frederic 1995a,b).  A generous linking length is preferred in studies
which aim to identify  high-velocity dispersion systems.  On the other
hand, studies of loose  associations require short velocity links, but
this   can  result  in   a  bias   towards  low   velocity  dispersion
measurements.   In   general,  the  velocity   dispersion  of  systems
identified with the FOF algorithm is $\sim 30 \% $ higher than
the  velocity  dispersion of  groups  identified  in  the same  galaxy
catalog by  the hierarchical method \citep{gar93,  giu01}.  To further
complicate  matters, clusters identified  with one  method may  not be
detected by the other.

Given the failings of traditional three-dimensional cluster
identification methods, which are likely to only be worse at high
redshift (due to evolutionary effects, sparse sampling etc.), we have
attempted to develop a new algorithm that may be applied to future
redshift surveys such as DEEP2.  This paper presents a new technique
for finding systems of galaxies from redshift space maps, designed
with the goal of matching their underlying, real space distribution.
Our algorithms for cluster detection and reconstruction use
three-dimensional Voronoi-Delaunay methods (hereafter VDM) to process
the spatial distribution of galaxies, which possess a number of
advantages over previous techniques.  We also present the results of
tests of these techniques using mock catalogs derived from N-body
simulations that mirror the properties of the DEEP2 survey.  Applying
our reconstruction scheme to artificial surveys, where the cluster
characteristics are known {\em a~priori} in real space, allows us to
clarify the uncertainties and assess the overall performance of the
method while at the same time testing the completeness of the
resulting catalog of systems.

The Voronoi partition of a space into minimally sized convex polytopes
-- the three-dimensional analogue of Dirichlet tessellation or
determination of Thiessen polygons \citep{dir50,vo08} -- provides a
natural way to find cluster centers (peaks in the galaxy density
field) without requiring any arbitrarily chosen window profiles or
smoothing length parameters.  The Delaunay complex \citep{del34}, the
simultaneously-determined dual of the Voronoi diagram, implicitly
contains vast amounts of proximity information and yields a natural
measurement of inter-galaxy scale lengths but is linearly proportional
in size to the dataset.  The Voronoi-Delaunay structures are a
fundamental tool of computational geometry and arise naturally and
independently in many different fields (see \citet{aur91} for a very
inclusive survey of Voronoi techniques in mathematics and the natural
sciences).  \footnote{In fact, what became known as the Voronoi
diagram was first suggested in an astronomical context for a problem
similar to that investigated here.  In his treatment of cosmic
fragmentation {\em Le monde, ou Le Traite de la Lumi\`ere},
posthumously published in 1664, R\'ene Descartes used Voronoi-like
methods to model the spatial distribution and the relative influence
of solar system bodies (see Fig.  7 in \citet{ok92}).}  More recently,
the Voronoi diagram has been applied in cosmological contexts to
investigate the problem of galaxy formation (e.g., 
\citep{ms84, iv87,  li87,  yi89,  vi89,  co91, it91}).  
Voronoi-based methods for selecting high-density regions in
two-dimensional images have also been developed (e.g.,
\citep{ebe93,pas94,ram01,kim01}).

Our cluster-finding technique is the logical extension of such
methods to the three-dimensional problem of finding clusters in a
redshift survey.  This VDM algorithm is intended to avoid some of the
difficulties which the standard hierarchical and percolation
algorithms present, particularly for large surveys
covering a wide redshift range.  It is a known fact that some cluster
parameters are more sensitive to the cluster definition procedure than
others.  For instance, \citet{mhg01} found
that the different global light distribution of systems, identified in
the Nearby Optical Galaxy (NOG) sample using different standard grouping methods, are
consistent with the hypothesis of being drawn from a common underlying
parent   distribution. Moreover, \citet{giu01}
show that the major dynamical properties of clusters (mass, virial
radius, etc.)  are relatively unstable, with values depending on how
the clusters and their members were identified using standard
clustering algorithms.  In this paper, we show that our method is optimized to
preserve the information encoded in key cluster
distribution functions, such as the number density of clusters as a
function of their line of sight velocity dispersion n($\sigma_{los}$)
and redshift n$(\sigma_{los}, z)$, even when $\sigma_{los}$
is as low as 200 \kms.

The outline of our paper is as follows: in \S 2, we summarize the
major technical characteristics of the DEEP2 Redshift Survey and
describe the mock catalogs used to simulate the survey volume and
galaxy fluxes. In \S 3 we present the group-finding algorithm and its
technical implementation; in \S 4 we examine the performance of this
algorithm when it is applied to the mock catalogs.  In \S 5 we apply
the DEEP2 target selection bias to the mock catalogs and derive the
characteristics and size of the expected DEEP2 optical cluster
sample. In \S 6 we investigate the relation between clusters
reconstructed applying the VDM to a flux-limited sample of halo
galaxies in redshift space to the underlying matter distribution
traced by fainter galaxies.  Our results are summarized in \S 7.

\section{The DEEP2 Redshift Survey}

Once observations commence  in the summer of 2002,  the DEEP2 Redshift
Survey  will be uniquely  capable of detecting  in three
dimensions and resolving the velocity structure of clusters and groups
of galaxies at high redshift, $z  \sim 1$.  A new instrument, the Keck
II Deep  Imaging Multi-Object Spectrograph (DEIMOS; cf. \citet{cow97}),  was designed for
this project  and will be installed  in Hawaii early  in 2002.  DEIMOS
can provide  imaging or multi-slit  spectroscopy over a field  of view
that is  approximately a rectangle  of size 16$\arcmin$  by 5$\arcmin$
and  over  the  wavelength  range  0.42-1.1  $\mu  m$.   
By  using  custom-milled
slitmasks, DEIMOS can  obtain spectra of $\sim 100-150$  galaxies at a
time. This  will allow  DEEP2 to observe  $\sim 60,000$  galaxies with
$0.7 < z <  1.2$ (a depth  which corresponds  to using  the  1200 lines/mm
grating and a magnitude limit $I_{AB} < 23.5$) over the 120 nights
allocated to the project, in addition to a deeper, smaller subsample.
The survey should be completed in late 2004.  All objects will be
observed with much higher spectral resolution than planned for other
comparable projects, $R \equiv \lambda / \Delta\lambda \sim5000$.  This
resolution will allow DEEP2 to measure redshifts of galaxies to a
precision of $<10$ \kms, readily resolving even small groups.

DEEP2 will target four fields each 2$^{\circ}$ by 0.5$^{\circ}$, which
corresponds to  comoving dimensions  of 80 by  20 $h^{-2}$  Mpc$^2$ at
$z\sim$   1   for   a   $\Lambda$CDM  universe.    The   corresponding
line of sight  comoving  distance   from  $z=0.7-1.2$  is  $\sim$1400
$h^{-1}$ Mpc, though as DEEP2  is flux-limited sampling will be sparser
at   the  highest  redshifts.    The  four   fields  were   chosen  in
low-extinction regions  which are continuously  observable from Hawaii
over  a six  month interval.   One field  includes the  extended Groth
Survey strip \citep{gro94},
and two fields
are on the  equatorial strip which will be deeply  imaged by the Sloan
Digital  Sky Survey  (SDSS).   Deep $BRI$  imaging  has been  obtained
throughout  each  field  using  the  CFHT  12k  camera  by Kaiser 
\& Luppino (see \citet{wil00} for details),
allowing   photometric   selection  of   objects   with  $z>0.7$   for
spectroscopy.   Many   observations  in  other   wavebands,  including
sensitive  Sunyaev-Zel'dovich studies  in each  field, have  also been
planned.

Because spectra  from adjoining  slitlets on the  same mask  cannot be
allowed  to overlap,  slitmask  spectroscopy is  inevitably unable  to
target every object in the  densest regions.  On average, spectra will
be  obtained  for  $\sim$70$\%$  of  all galaxies  meeting  the  DEEP2
selection criteria  in each  field, but sampling  will be  lower where
galaxies are most densely packed  on the sky.  Coil, Davis, \& Szapudi
(2001, hereafter  CDS) tested the degree to which  this selection  bias may
affect  measurements  of  the  underlying  two-point  and  three-point
correlation functions.   For that purpose, they  created mock catalogs
which emulate the  properties of a flux-limited DEEP2  sample; we have
made use of those same catalogs to test the performance and robustness
of our cluster-finding algorithm.

These mock data were constructed from simulations that combined the
results of high-resolution, large-volume N-body calculations with
semi-analytic methods that model the formation and evolution of
individual objects \cite{kau99}.  In particular, the catalogs provided
by Kauffmann et al. were based upon the GIF N-body
simulations\footnote{http://www.mpa-garching.mpg.de/Virgo}; CDS used
the $\Lambda$CDM catalogs, based on a model with
$\Omega_{matter}$=0.3, $\Omega_{\Lambda}$=0.7, $h$=0.7, and
$\sigma_8$=0.9.  To match the DEEP2 survey, CDS then constructed from
the Kauffmann et al. results six nearly-independent volumes with the
geometry of a DEEP2 field, each covering the equivalent of 2$^{\circ}$
by 0.5$^{\circ}$ on the sky and a redshift range of $0.7<z<1.2$.  At
this point,volume-limited samples could be assembled; however, DEEP2
will be a flux-limited survey.  Coil, Davis, \& Szapudi thus used the
absolute magnitudes for each galaxy provided by the semi-analytic
calculations to select all objects brighter than the DEEP2 magnitude
limit, resulting in six mock catalogs containing $\sim 15,000$
galaxies each.  The robustness of the VDM is tested using all six
DEEP2 mock catalogs.

\section{Cluster finding method}

It is difficult to develop  a single method that can robustly identify
and determine the membership of groups 

\vbox{%
\begin{center}
\leavevmode
\hbox{%
\epsfxsize=15.9cm
\epsffile{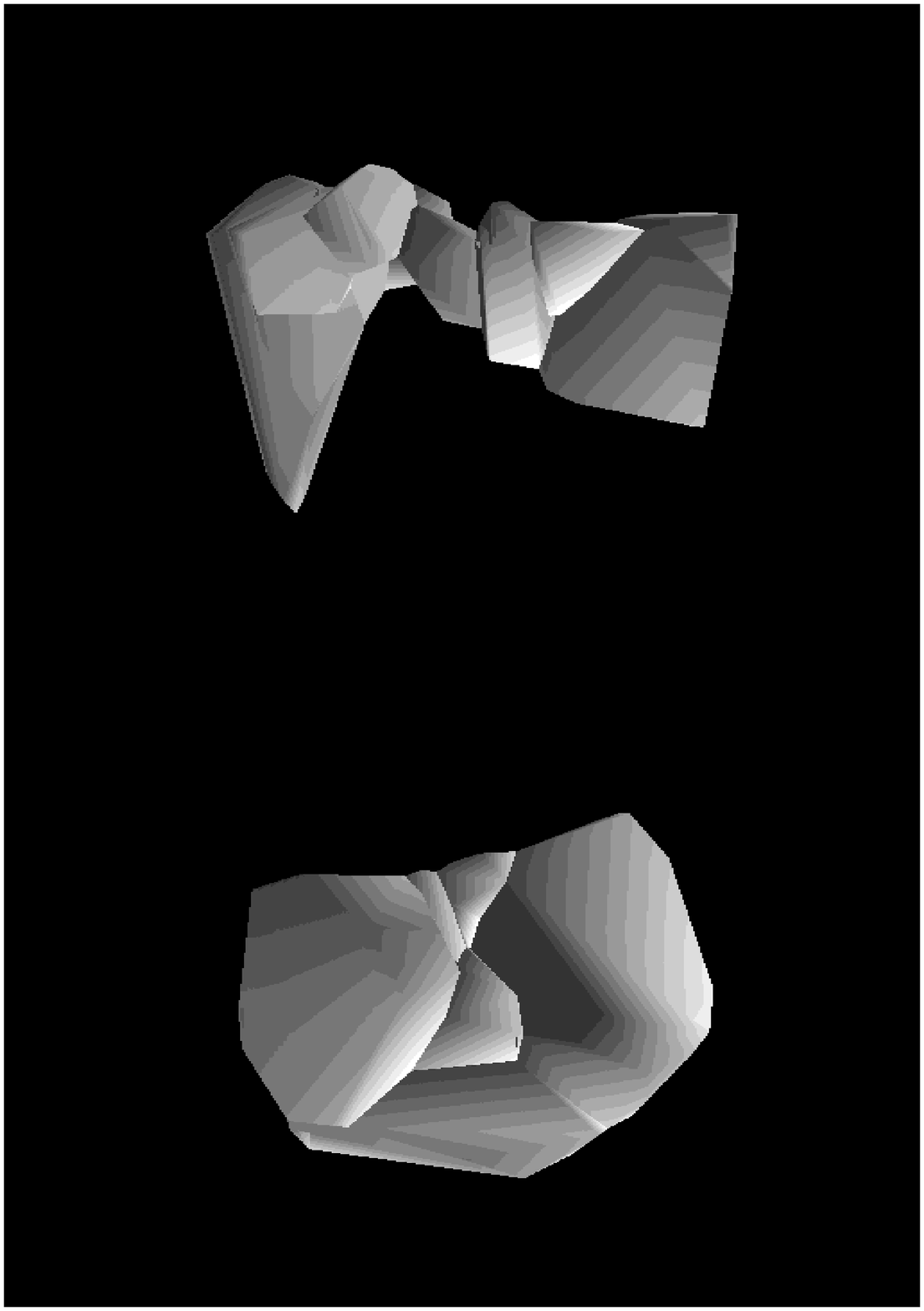}}
\begin{small}
\figcaption{%
Three-dimensional Voronoi reconstruction of a cluster with 10 galaxies
in the DEEP2 mock catalog. The Voronoi cells encompassing each cluster
galaxy are shown in real space ({\em bottom}) and in redshift space
({\em top}).  Each Voronoi 3D cell surrounding a galaxy is defined as
the intersection of the planes which are perpendicular bisectors of
the lines joining that galaxy to the others. Note how the isotropic
real-space distribution of cluster galaxies degenerates into a
composite Voronoi structure which is elongated along the observer's
line of sight.
\label{fig1}}
\end{small}
\end{center}}
\clearpage

\vbox{%
\begin{center}
\leavevmode
\hbox{%
\epsfxsize=8.9cm
\epsffile{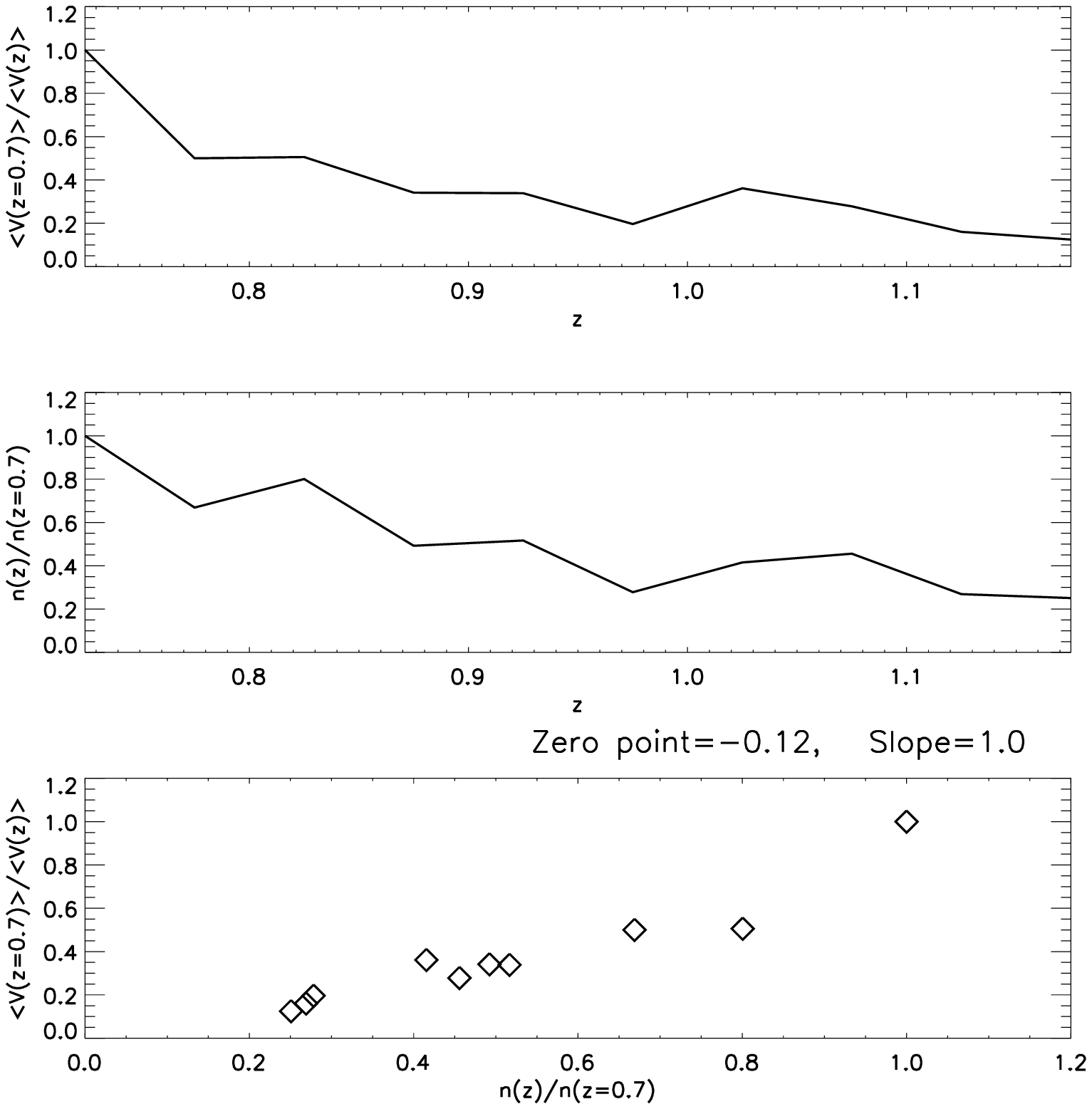}}
\begin{small}
\figcaption{%
The mean galaxy density evaluated by averaging over Voronoi 3D cells contained
in a volume $\Delta$V(dz=0.05) ({\em upper}), and by averaging over the counts of 
galaxies (corrected for relativistic effects) in the same region $\Delta V$
({\em center}). 
{\em Lower:} a comparison between the two estimates shows the consistency. 
%CCT: UPDATE THIS IF THE PLOT CHANGED
\label{fig2}}
\end{small}
\end{center}}

\noindent  and clusters of galaxies across
a wide range of masses and redshifts.  These structures may range from
$\sim10^{12}M_{\odot}$  to  $\sim10^{15}M_{\odot}$  in mass  (with B luminosity
varying from $\sim10^{10}L_{\odot}$  to $\sim10^{13}L_{\odot}$). Moreover, since
the mass density of a cluster should depend primarily on its formation
time (in the spherical-collapse  paradigm), some small groups and rich
clusters may  have the same  mean separation between  member galaxies.
As a  further complication,  flux-limited surveys cannot  have uniform
sampling at all distances.  As a result, two groups that have the same
physical  characteristics  (mass,   size,  richness,  etc.)  may  have
different populations  of galaxies observed  if they are  at different
redshifts, even without galaxy evolution effects.

We  have  attempted  to  overcome   these  problems,  at  least  in  a
statistical sense,  by taking  a computational approach  that utilizes
both density  and spatial  proximity information to  identify clusters
and their members.   The Voronoi diagram and Delaunay complex  form the basis
for    this   technique.    The    Voronoi   partition    provides   a
geometrically-defined measure  of the local  density field of  a point
process, while  the Delaunay triangulation  contains information on
the  spatial relationships between  those points.   We will  show that
using these tools,  we may objectively find clusters  and that their
intrinsic properties can be reliably measured.

\subsection{The Voronoi diagram and Delaunay complex}

A Voronoi polyhedron is the uniquely defined convex region of space
around a chosen object (also referred to as the ``seed'') within which
each point is closer to the seed than to any other object.  The faces
of the Voronoi cell are formed by planes perpendicular to the vectors
between the seed and its neighbors (see figures \ref{fig1} and
\ref{fig3}).  The Voronoi partition of a space into minimally sized
convex polytopes provides a natural way to measure packing.  The
volume inside each polyhedron is inversely proportional to the packing
efficiency of its seed; a large cell volume indicates that its seed is
comparatively isolated.

By performing  a Voronoi decomposition  on a redshift catalog,  we may
estimate   the   three-dimensional   galaxy   density   field   in   a
non-parametric way.   Other methods  estimate this field  by smoothing
the distribution of data points with an {\em a~priori} physical model,
window profile or binning strategy \citep{ebe93, mhg01}.  In contrast,
the  Voronoi diagram provides  a density  estimator  that is
asymptotically local \citep{fad97}; the density measured at a position
${\bf x}$ is determined completely by the positions of the neighboring
data points, while the influence of distant points vanishes.

The Delaunay complex in three-dimensional space is defined by the
tetrahedra whose vertices are sets of four galaxies with the property
that the uniquely determined sphere circumscribing them does not
contain any other galaxy.  The center of this sphere is the vertex of
a Voronoi polyhedron.  The Delaunay triangulation represents the
geometrical dual of the Voronoi partition and provides a natural
linking structure for a set of objects.  A complete treatment of the main
mathematical characteristics of these geometrical structures can be found 
in \citet{zan90} or \citet{vdw94}.

We  calculate the  Voronoi  and Delaunay  structures  as described  in
\citet{bdh96}  and compute  moments  over each  Voronoi cell  (volume,
center of mass, and moment  of inertia) following the prescriptions of
\citet{mir96} to transform volume integrals within a polyhedron into
explicit sums over its vertices.   As an example, in figure \ref{fig2}
we show that the mean  density obtained by averaging over the ensemble
of Voronoi cells, $\langle 1/V\sbr{i} \rangle\sbr{V}$, agrees with the
standard density  estimator $N/V$,  where $N$ is  the total  number of
objects and $V$ is the volume over which the average is computed.

\subsection{Details of the algorithm}

Once  the   Voronoi/Delaunay  calculation  for  a   catalog of galaxies 
in the Doppler distorted redshift space  (we know real positions 
and peculiar velocities for each galaxy in the catalog)
has  been completed,  our algorithm  proceeds  in three  phases.  First,  global
minima of  the Voronoi  cell volume (i.e.,  peaks in the  density) are
identified and  provide candidate locations for  cluster centers.  The
Delaunay  mesh then  allows us  to  identify central  members of  each
candidate  group   and  estimate  physical  properties such as the cluster
central density. Finally, these  estimates  are used  to
define  redshift space  windows  within  which we  find  each  group's
members.  In this  last step it is the predicted structure of the clusters, 
inferred  from an initial level  of grouping, that influences {\em local}
decisions regarding galaxy membership.

Before discussing each step in more detail, we state first the working
definition of a cluster we have used to judge the performance of our
grouping method. 
\vbox{%
\begin{center}
\leavevmode
\hbox{%
\epsfxsize=8.9cm
\epsffile{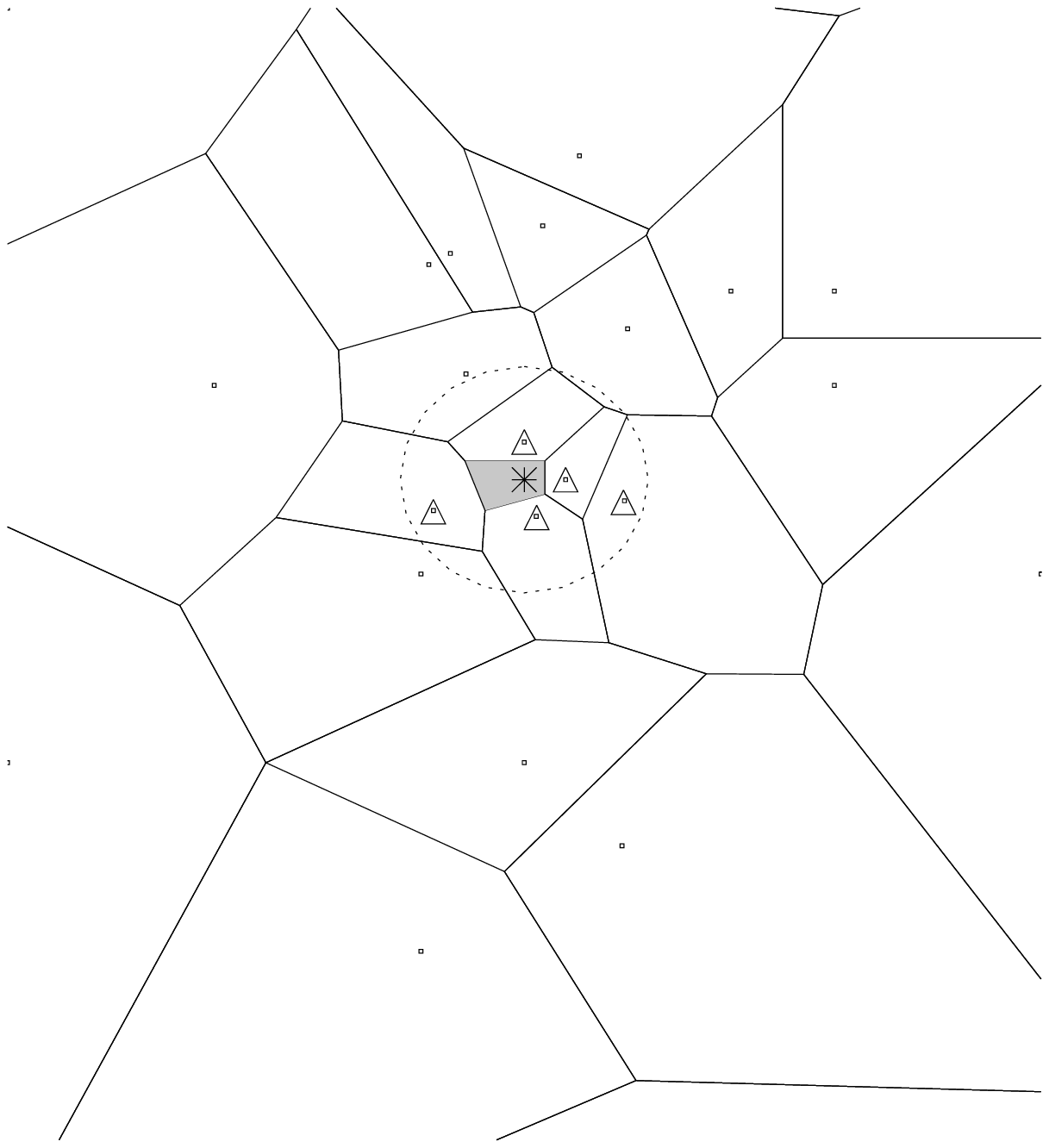}}
\begin{small}
\figcaption{%
Simplified 2-dimensional  graphic representation of phase I (\S 3.2.1).
Sky angular coordinates are along the x axis, and the survey 
depth is displayed  along the y axis. 
Dots represent  the galaxy distribution, while
the shaded area represents the Voronoi cell surrounding a possible cluster seed
(represented by an asterisk). The set formed by the asterisk and the 
points marked with triangles represents the
{\em first order Delaunay neighbors}.
\label{fig3}}
\end{small}
\end{center}}

 We consider a cluster to be a system identified as a
single group in the real-space, volume-limited DEEP2 mock catalogs
using a FOF algorithm with a linking length parameter $b=0.2$.  This
parameter value has been shown to effectively select virialized
overdensities (mean overdensity $\sim$ 180 in a critical universe), at
least in the statistical sense \citep{cl96}, and it is widely used by
simulators in deriving the mass statistics for their N-body halos (e.g.
\citet{ben00} or \citet{jk00}); this choice, therefore,
facilitates comparison to the literature.

Although this serves to define the set of clusters and their
membership in each mock catalog, it is not always desirable to compare
a given set of VDM results to this ``reality.''  For instance, 
to assess the performance of our method, i.e.  
determining what fraction of group members are recovered in the same
VDM group, we must compare not to the original group membership, but
rather only to those members of the actual group which are brighter
than the DEEP2 flux limit, as they alone are present in the catalogs
to which we apply our algorithm.  Except where otherwise specified
(e.g. \S 6), in
the remainder of this paper we compare VDM results to the real-space
friends-of-friends catalog {\em after} flux selection; e.g., in $\S 4$
we compare the velocity dispersion determined for each VDM cluster to
the dispersion calculated from those members of that cluster's
counterpart identified in real space that are brighter than the DEEP2
magnitude limit.  Any biases in the DEEP2 VDM sample should then
similarly affect the comparison sample, and we can test how well our
algorithm reconstructs the information present in the redshift
catalog.

\vbox{%
\begin{center}
\leavevmode
\hbox{%
\epsfxsize=8.9cm
\epsffile{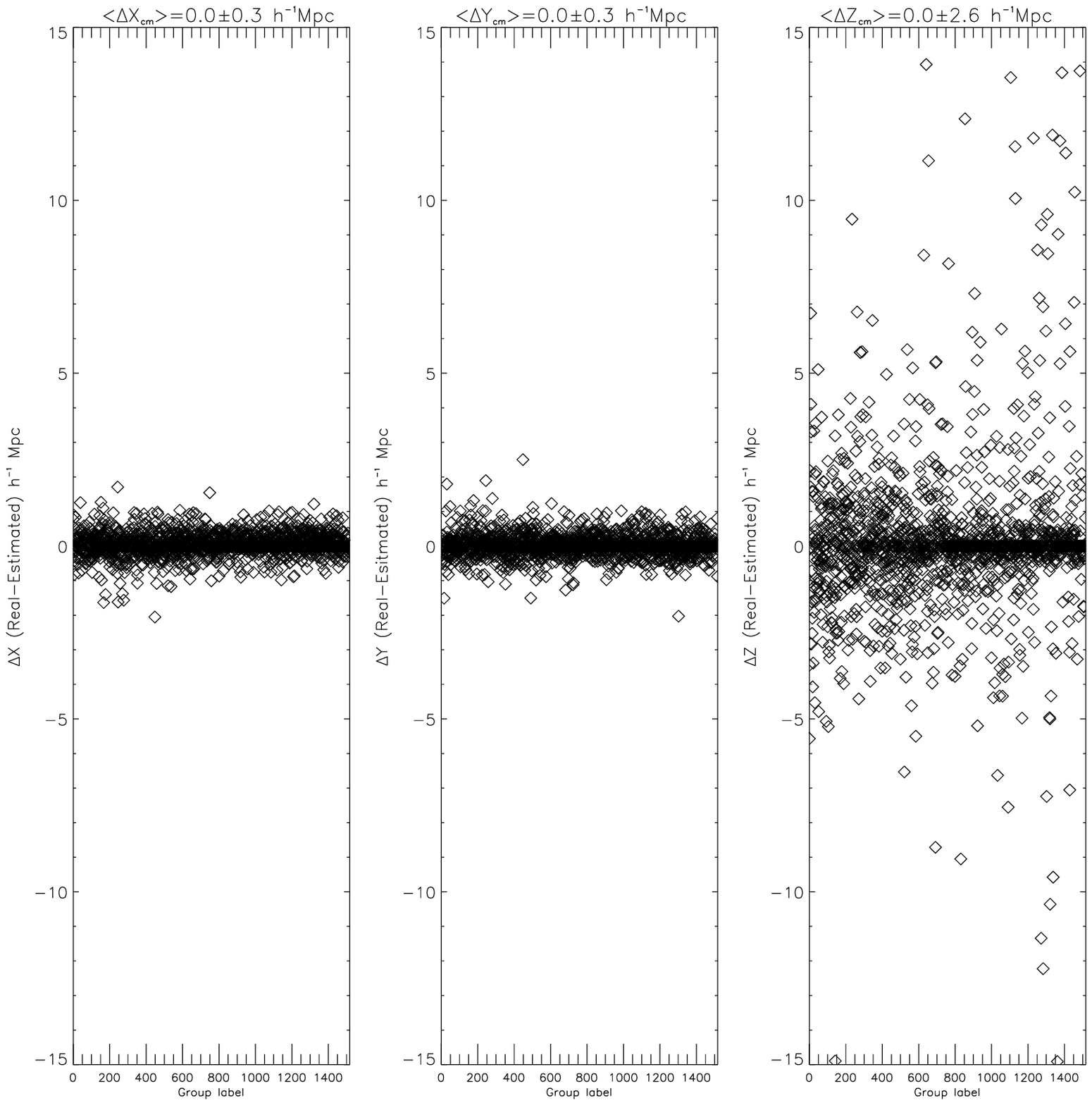}}
\begin{small}
\figcaption{%
Deviations between the three Cartesian components of the
estimated group  barycenter positions  (from the 
{\em first-order Delaunay neighbors}) and the center of mass 
position of the underlying real group.
The underlying real structure is computed by determining which 
members contributing to $N\sbr{II}$ belong to the same structure in real
space and using  as a reference the biggest of these real systems.
Groups are labeled in order of increasing richness along the x axis.
On the top of the panel the mean and the  1$\sigma$ deviations
are reported.
\label{fig4}}
\end{small}
\end{center}}

\subsubsection{Phase I: Finding systems of galaxies}

We begin by assuming that the centers of clusters will lie near peaks
in the galaxy density field (this assumption is tested below).  To
identify these peaks, we sort all the galaxies in the catalog by the
inverse volume of their Voronoi cell; the smallest cells are most
likely to fall at density maxima and are thus potential ``seeds'' for
finding groups or clusters.  We must next determine if a given seed
actually lies at the heart of a system of galaxies.

Each  cluster seed  will be  linked to  its nearest  neighbors  by the
Delaunay mesh.  We are interested  only in real, physical groupings of
galaxies; we therefore  must define some {\em ad  hoc} threshold in an
attempt  to  distinguish  galaxies  that  could  be  physically  bound
together from galaxies which are in chance proximity to each other.
We  consider to  be  neighbors those  Delaunay-connected points  whose
distance from  the seed  galaxy is less  than a fixed  limit ${\cal
R}\sbr{min}$.  These  galaxies, and  the original seed  galaxy itself,
will be referred to  hereafter as {\em first-order Delaunay neighbors}
and are used to determine the system's center of mass.

For  this  paper we take ${\cal  R}\sbr{min}=1$ \hmpc~comoving as standard. 
In a survey such as DEEP2 that selects galaxies down
to $\sim  L^*$, a typical loose  system like the Local  Group would be
detected with only 2 elements $\sim 1$ \hmpc ~apart.  
\vbox{%
\begin{center}
\leavevmode
\hbox{%
\epsfxsize=8.9cm
\epsffile{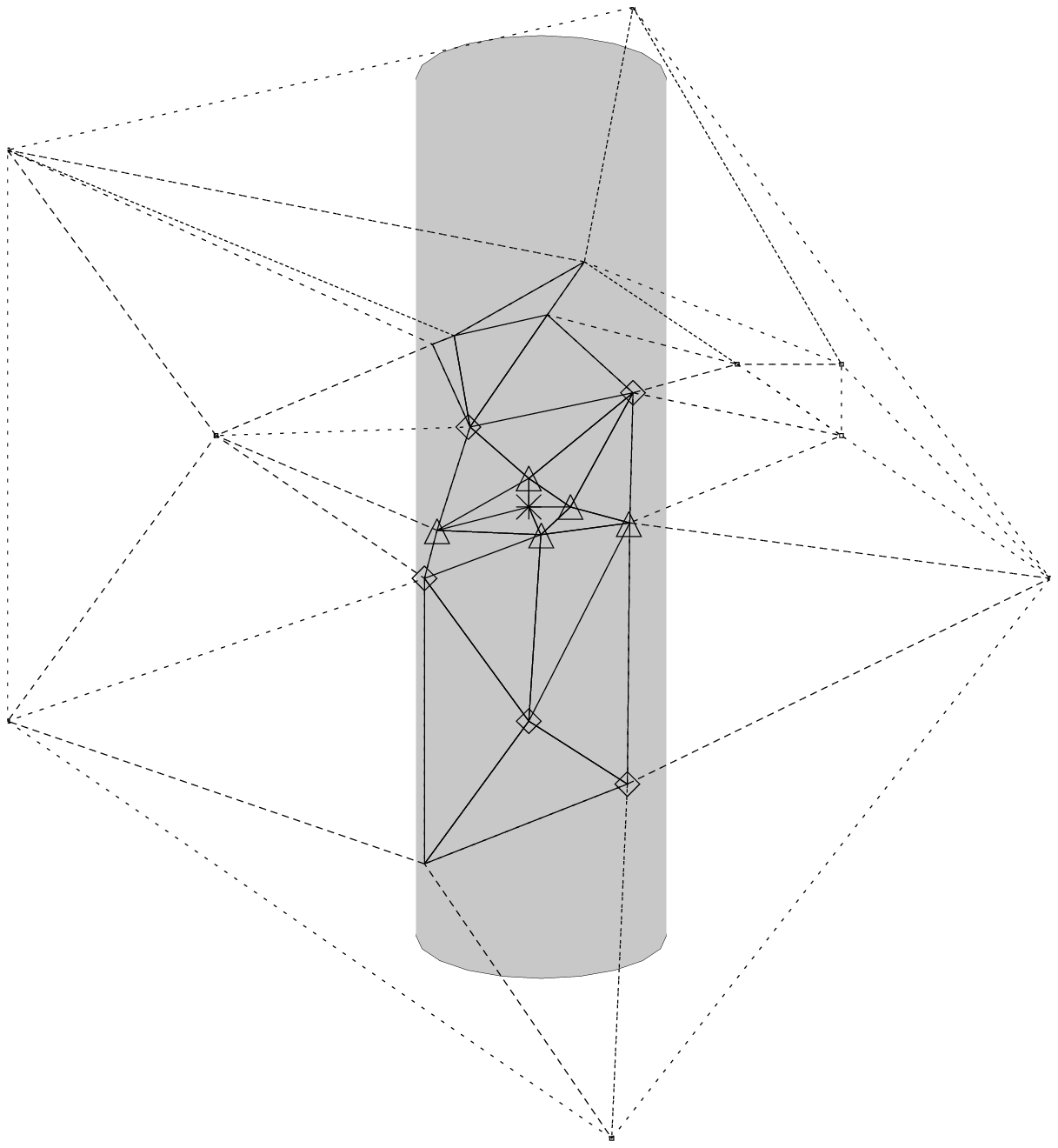}}
\begin{small}
\figcaption{%
Simplified 2-dimensional graphic representation of phase II (\S 3.2.2). 
Sky angular coordinates are along the x axis, and the survey 
depth is displayed along the y axis.
Segments represent the Delaunay mesh 
connecting the galaxy distribution of figure \ref{fig3}. 
{\em Second-order Delaunay neighbors} are represented by diamonds.
Note that not all the galaxies inside the search window (shaded area) are used to define the projected 
central density parameter $N\sbr{II}$, but only those designated by symbols.
\label{fig5}}
\end{small}
\end{center}}
This is also the
typical  mean  projected separation  between  pairs  of galaxies  within
clusters found via a  FOF algorithm \citep{mariphd}, and it is
the typical central  radius of massive clusters as  used in studies of
the  correlation  between central  richness  and velocity  dispersions
\citep{bah81}.

If no  galaxies satisfy this criterion  then the cluster  seed will be
rejected and considered an isolated galaxy.  If, when analyzing a seed
we find that all its {\em first-order Delaunay neighbors} have already
been assigned  to another structure,  we automatically merge  the seed
galaxy  into that system.   A schematic  representation of  this first
step in cluster identification is presented in figure \ref{fig3}.

The search radius ${\cal R}\sbr{min}$, which determines if a point
must be considered isolated or not, primarily serves as a parameter
controlling the final number of identified systems. The dependence of
the cluster statistics on this parameter will be investigated in \S 4.
In this first phase, the search radius allows us to exclude the most
deviant points from our center of mass estimates.

Adjusting ${\cal R}\sbr{min}$ within a range $\pm50\%$ has minimal
effects on the derived cluster center positions.  In fig. \ref{fig4}
we compare the cluster center of mass recovered in redshift space with
the center of mass of the counterpart structure identified in real
space.  All three spatial components, when averaged over the whole set
of mock catalogs, show an offset of zero and a standard deviation less
than 0.3 \hmpc ~and 2.5 \hmpc ~in projection and along the line of
sight respectively, testifying to the effectiveness of this technique.

\subsubsection{Phase II: Determining clustering strength} 

Since they are calculated in a parameter-free fashion, both the
Voronoi diagram and Delaunay complex are determined isotropically in
the angular and redshift directions.  However, the peculiar velocities
induced by a cluster's gravitational field cause the distribution of
galaxies to appear elongated in the redshift direction to a degree
determined by its velocity dispersion.  The three-dimensional
information lost in the transformation to redshift space cannot be
recovered uniquely via isotropic, geometric methods; additional
assumptions are required to minimize contamination by spurious
members.  To determine the properties of clusters with any accuracy,
we require methods that include this anisotropy.

We therefore  define a cylindrical window in  redshift space (centered
on the  barycenter determined in Phase  I and circular  in the angular
dimensions) within which we may  find objects which are very likely to
be members  of each  cluster.  This cylinder  will have  radius ${\cal
R}\sbr{II}  \geq {\cal R}\sbr{min}  $ and  length (along  the redshift
direction) ${\cal  L}\sbr{II}$.  We  define those galaxies  which fall
within  this  cylinder  and  are  connected  to  first-order  Delaunay
neighbors  by   the  Delaunay  mesh  as   {\em  second-order  Delaunay
neighbors}; see figure \ref{fig5} for a graphical illustration. Note 
that not all the galaxies in the cylinder are included.

We set ${\cal R}\sbr{II}= {\cal R}\sbr{min}=1$ \hmpc~comoving, which
is the typical central radius of massive clusters as used in studies
of the correlation between central richness and velocity dispersions
\citep{bah81}.  Analogous physical considerations guide us to set the
half-length of the cylindrical window, ${\cal L}\sbr{II}$, to be 20
\hmpc. This value includes the upper limit of the peculiar velocity of
galaxies that are members of real systems in our simulations (as high
as $\sim 2000$ \kms; cf. \ref{fig7}), taking into
account the relativistic stretch factor of $\sim 2$ in peculiar velocities at
redshift z$\sim$1 (see Appendix A) and the error in the precision with
which we can fix the cluster barycenter (see $\S$ 3.2.1).

We may use the sum of the numbers of {\em first- and second-order
Delaunay neighbors} as an indicator of the central richness of the
group, $N\sbr{II}$.  This parameter ranges from 2 to 25 for the
clusters identified in the DEEP2 mock catalogs (note that the cluster seed is
included).  Since the window has fixed volume,
$N\sbr{II}$ corresponds also to an estimate of the central density of
galaxies; for less massive systems with small velocity dispersions,
it should also be roughly proportional to the projected surface
density as measured from images.  By including only Delaunay
neighbors in $N\sbr{II}$, we are able to minimize contamination by
interlopers, providing a robust estimate even in low-density systems.
This is particularly important because $N\sbr{II}$ controls the
adaptive window for cluster members used in Phase III.

\subsubsection{Phase III: Assigning cluster members} 

Having detected the center and estimated the richness for each
cluster, we then reconstruct the full set of system members.  We do
this on the basis of physical considerations, not via an empirically
tuned parameter threshold.  In particular, we exploit the known
richness-velocity dispersion correlation to define a search window for
each cluster's members based upon its richness.

\vbox{%
\begin{center}
\leavevmode
\hbox{%
\epsfxsize=8.9cm
\epsffile{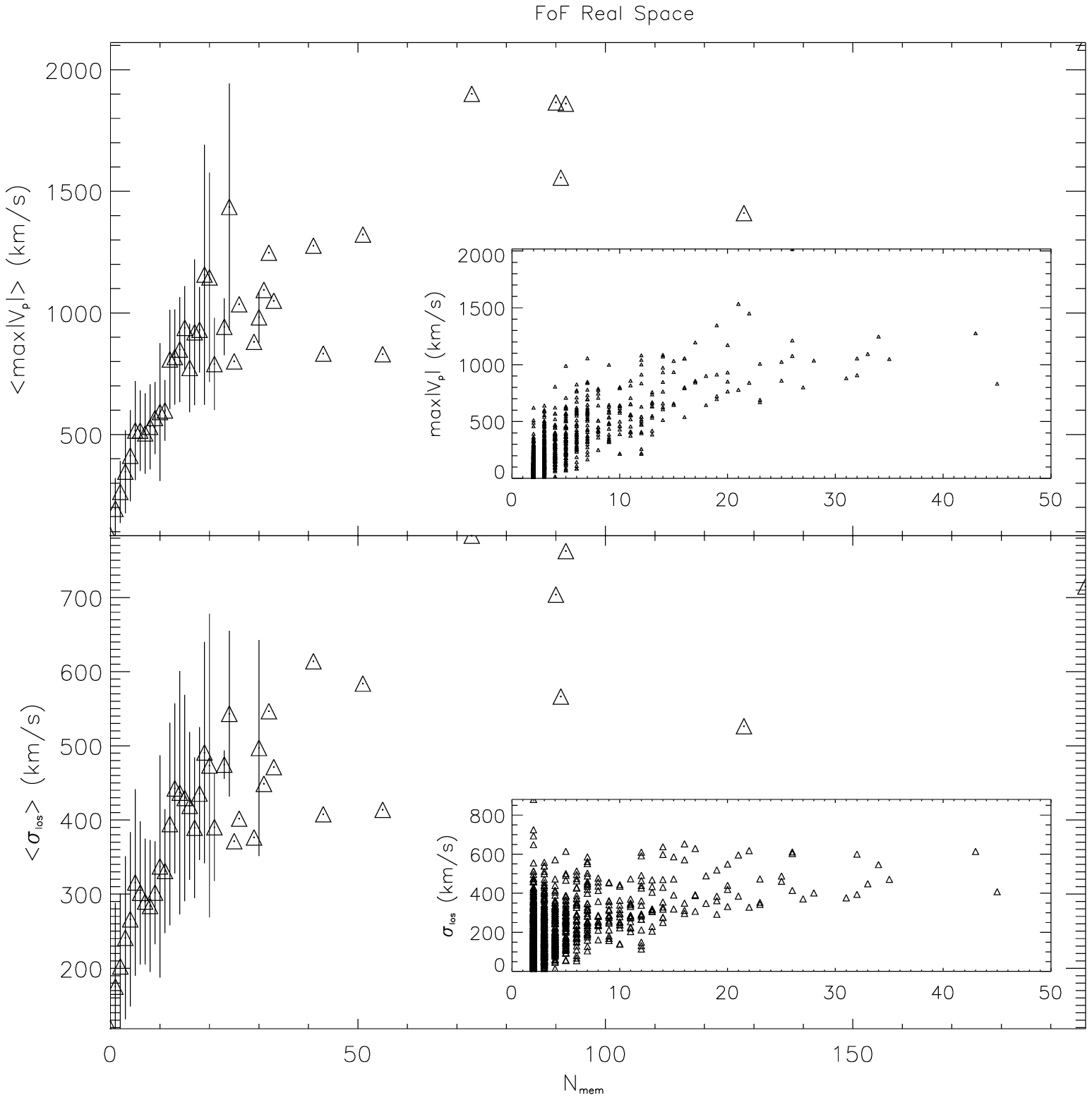}}
\begin{small}
\figcaption{%
{\em Upper:} the mean of the maximum of the absolute value of the peculiar 
velocities of real-space identified cluster members, calculated with respect to the cluster barycenter,
is shown as a function of the cluster richness (for mock catalog \#1).
Errorbars represent the 1$\sigma$ standard deviation of the mean.
The inset shows the scatter plot of max$|v_p|$ as a function of the cluster 
richness over the lower portion of the richness range.
 {\em Lower:} the mean line of sight velocity dispersion is plotted as a function
of richness for real-space clusters. The inset shows the scatter plot of the line of sight 
velocity dispersions as a function of 
the cluster richness over the lower portion of the richness range.
\label{fig7}}
\end{small}
\end{center}}

The virial theorem predicts that velocity dispersion and central
number density of galaxies are correlated.  \citet{bah81}
observationally confirmed the existence of a strong linear correlation
valid from loose groups to clusters. Such a relation in fact holds for
the clusters found in our mock catalogs; in figure \ref{fig6} we show
that a linear trend is seen when we plot the maximum peculiar velocity
of galaxies in a system (here represented as a distance stretch)
versus the cluster central richness.  We rely on this relation to
estimate the strength of the underlying clustering, which we then use
to determine on a group by group basis the window around the system's
center within which to search for Delaunay-connected galaxies.

Specifically, for each cluster we define a cylindrical window
(symmetric about the redshift direction) with dimensions determined by
the richness scale factor $N\sbr{II}$ and then process the set of
Delaunay-connected galaxies inside with a rapidly converging
``inclusion-exclusion'' logic to identify cluster members within that
window. By using the Delaunay mesh to identify the nearby galaxies, we
are able to do this quite rapidly; once the initial Voronoi-Delaunay
calculation is complete (which need only be done once for a catalog),
it takes only 5 minutes on a modern workstation to process $\sim
19000$ galaxies into a catalog of groups and clusters.

\vbox{%
\begin{center}
\leavevmode
\hbox{%
\epsfxsize=8.9cm
\epsffile{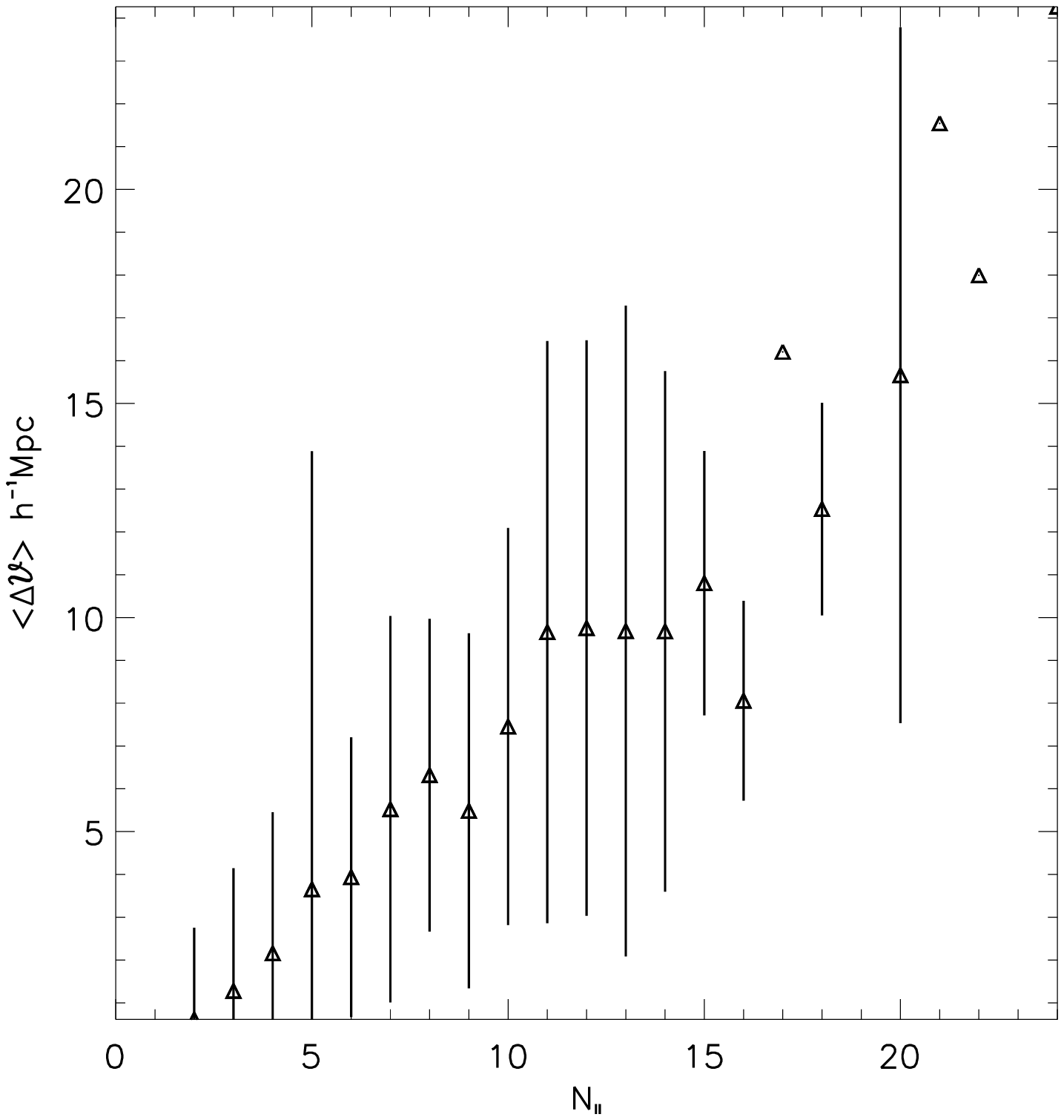}}
\begin{small}
\figcaption{%
Mean line of sight half-length dimensions of real groups
as measured in  redshift space
are plotted versus the central density parameter $N\sbr{II}$ defined in section 
3.2.2. The measured elongation is caused by galaxy peculiar velocities. 
\label{fig6}}
\end{small}
\end{center}}

The radius of the window (in the plane of the sky) and its half-length
(in the redshift direction) are determined by the equations:
\begin{equation}
\left\{
\begin{array}{l}
{\cal R} = \max [r \cdot  N\sbr{II}, {\cal R}\sbr{II} ] \\ {\cal L} =
\max [v \cdot N\sbr{II}, {\cal L}\sbr{min} ]
\end{array}
\right.
\label{sys}
\end{equation}
where $r$ and $v$ are coefficients that fix the scale of the window in
each direction and ${\cal L}\sbr{min}$ is a cutoff filter required to
take into account the loss of distance resolution on small scales from
the smearing caused by peculiar velocities.

We  have determined  $r$ and  $v$ by  requiring that  when $N\sbr{II}$
takes  its maximum possible  value, the  window has  dimensions ${\cal
R}_{max}$ and  ${\cal L}_{max}$ that correspond to  the maximum radius
and peculiar velocity expected for a cluster in a DEEP2 field.  Guided
by our  mock catalogs and literature  data \citep{abe58,bah89, bggg97}
we   assume  that   such  a   cluster  is   characterized   by  ${\cal
R}\sbr{max}\sim   1.5$   \hmpc~comoving    and   ${\cal   L}\sbr{max}=   {\cal
L}\sbr{II}=20$ \hmpc  ~(we expect $\sim 1$ Coma-like  cluster with 1000
\kms  ~dispersion   in  each   DEEP2  field  of   volume  $\gtrsim10^6$
\hinvmpcub).    ${\cal  R}_{max}$  then   matches  the   Abell  radius
\citep{abe58} while  the value of ${\cal L}\sbr{max}$ has already been justified
in the previous section. In a  similar way, the  threshold filter
${\cal L}\sbr{min}=5$  \hmpc ~is set by  noting that it is common to have
at  least  one  member with  a  peculiar
velocity of 500 \kms ~with respect to the cluster barycenter, even in systems 
of very low richness (see figure \ref{fig7}).
Although plausible physical considerations  have led to our definition
of the window size, it may also be justified {\em a~posteriori} by the
good agreement of the derived cluster membership with that found by FoF
in real space (see  $\S 4$).
\vbox{%
\begin{center}
\leavevmode
\hbox{%
\epsfxsize=8.9cm
\epsffile{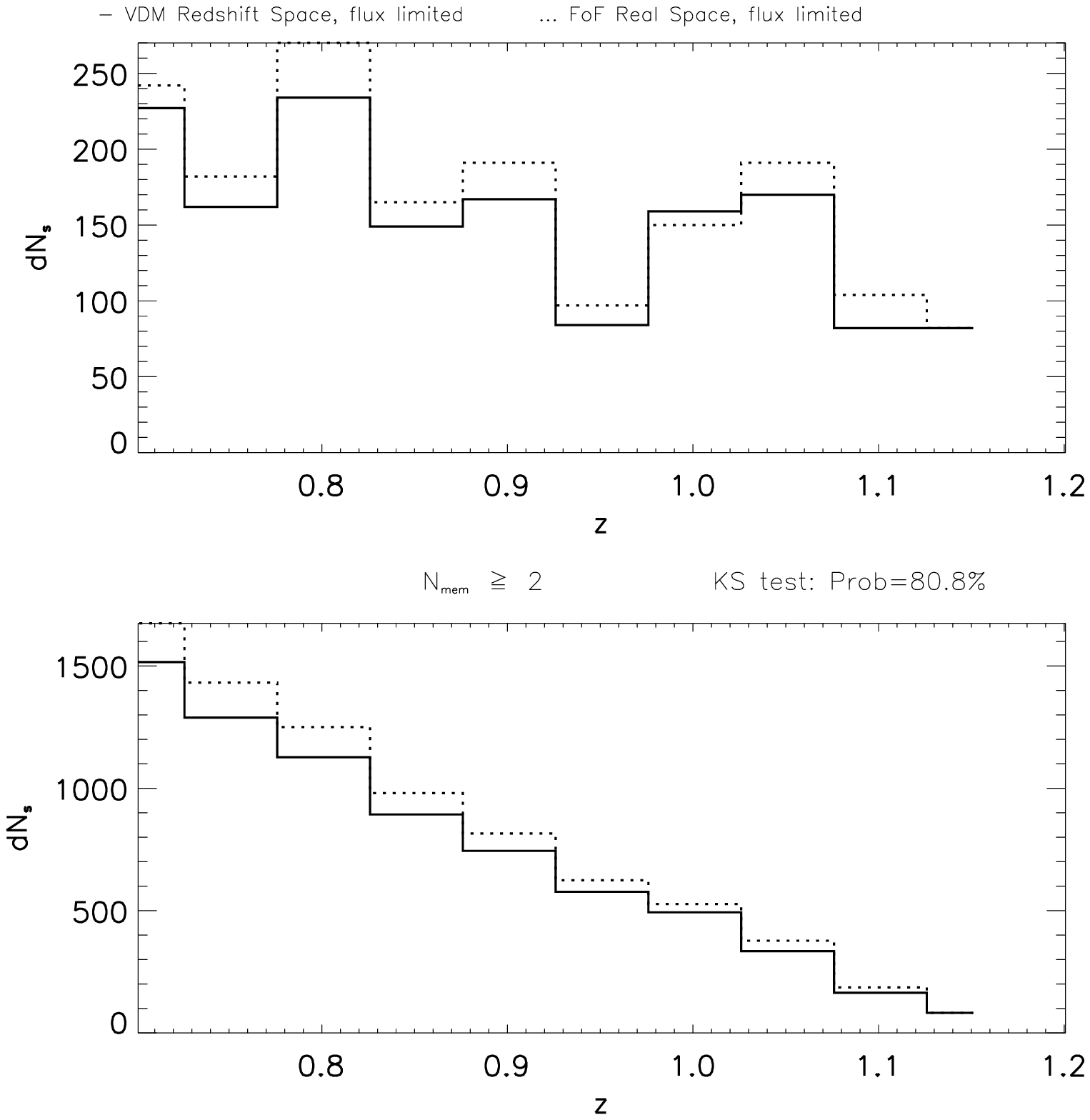}}
\begin{small}
\figcaption{%
Differential ({\em upper}) and cumulative ({\em lower}) distributions
of clusters identified by VDM in mock \#1 (whose galaxy distribution
is shown in figure \ref{fig10} and \ref{fig11}) as a function of
redshift.  The parent distribution (clusters identified in real space)
is the dotted line, while the observed distribution recovered by VDM
in redshift space is represented by the solid line. Data are
binned in redshift intervals of width 0.05.  A Kolmogorov-Smirnov
statistical test confirms the consistency and the lack of any
identification bias.
\label{fig8}}
\end{small}
\end{center}}

\vbox{%
\begin{center}
\leavevmode
\hbox{%
\epsfxsize=8.8cm
\epsffile{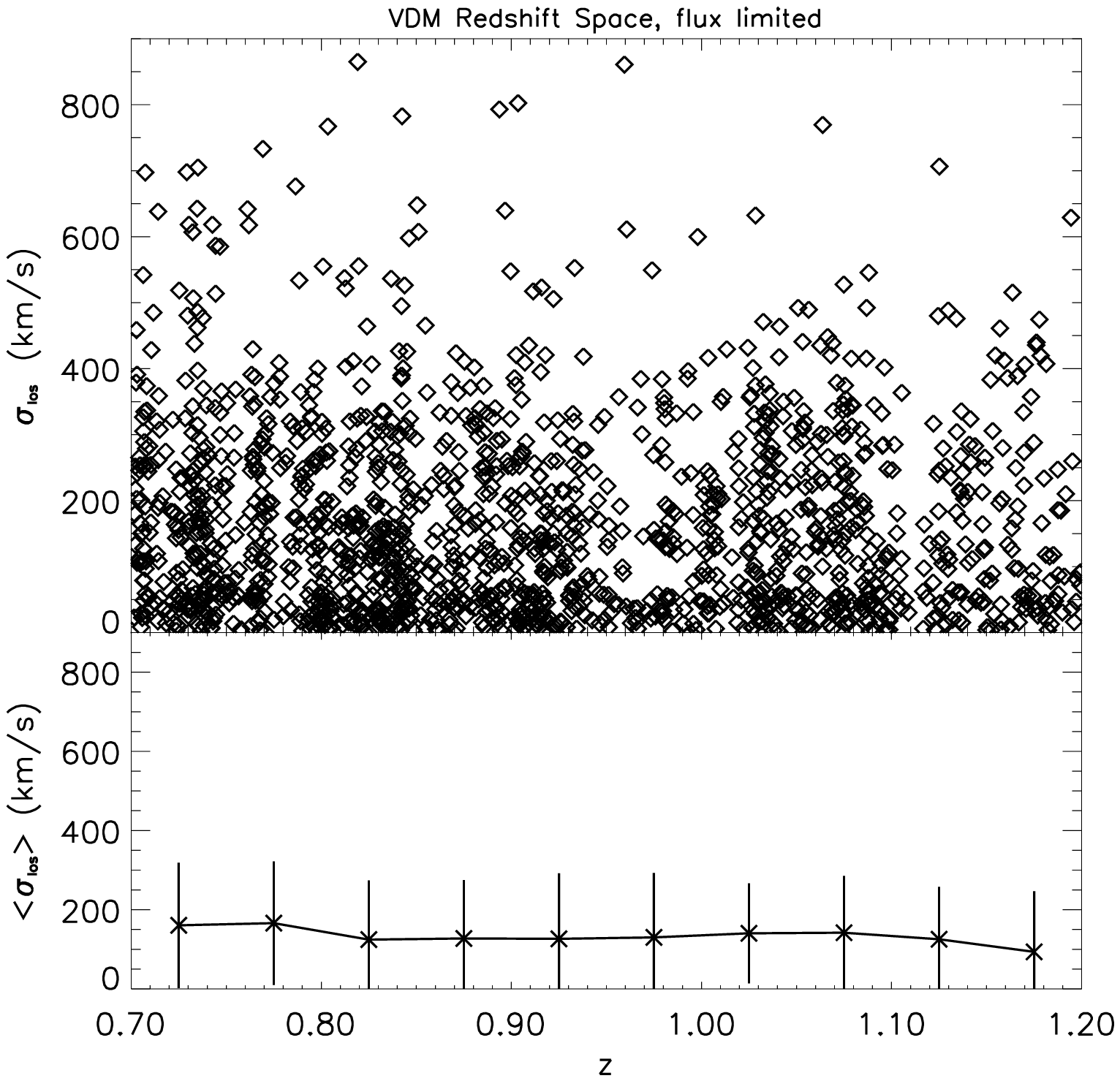}}
\begin{small}
\figcaption{%
{\em Upper:} velocity dispersions of the 
reconstructed groups  ($N\sbr{mem} \geq 2$)
in the mock \#1 catalog 
are plotted versus the survey depth.
{\em Lower:} the same data
averaged in redshift bins of width 0.05. Errorbars represent the standard deviation of the velocity dispersions of all systems in each bin.
\label{fig9}}
\end{small}
\end{center}}

Since the size of the search window depends upon the richness estimate
of  phase II,  to  ensure a  uniform  population of  clusters we  must
correct that estimate for the  variation of the DEEP2 luminosity limit
with redshift.  We do this based  upon the scaling of the Voronoi cell
volume as a function of redshift; as the sampling decreases at greater
distance, the density  of objects in the sample  will become lower and
thus  the mean volume  of the  cells larger.   We therefore  correct a
value of $N\sbr{II}$ at redshift $z$ to
\begin{equation}
N\sbr{II,corr}=N\sbr{II}                                          \cdot
\Big(\frac{<V(z)>}{<V(z=0.7)>}\Big)^{a},
\end{equation}

\noindent where $V$  is the volume of a  Voronoi cell and $a$
is   a  free   parameter  we   fit   by  comparing   the  scaling with 
distance of the reconstructed number of clusters with the analogous  
quantity calculated in the volume-limited real space simulation (see
figure \ref{fig8}.)

Although one might expect $a=1/3$ from simple scaling arguments, for
the mock DEEP2 catalogs we find $a$ is consistent with being $\lesssim
0.1$; i.e., the sizes of the windows need be only weakly dependent on
the galaxy density gradient.  There are two reasons for this: first,
in the {\em identification} phase, the Voronoi partition highlights
regions of enhanced clustering, where galaxies tend to be more
luminous \citep{park94, giu01}.  This can be appreciated in figure
\ref{fig8} where we plot the distribution of the number of identified
systems as a function of distance and show the lack of any
identification bias.  Second, the centrally static logic of our {\em
reconstruction} phase bypasses the problem common to other methods
(which usually link successive elements only relative to the last
merged unit) of breaking a distant system into sub-units as soon as
faint linking elements disappear under the visibility threshold of the
survey.
% and cannot be used as a linking unit. 
This can be inferred in figure \ref{fig9} where no systematic
correlation of the recovered velocity dispersion with survey depth is
apparent.

%CCT: I'm actually a little bit worried by the preceding.  We expect some evolution in the velocity function with redshift, though in this sample it would only occur in regions the size of the survey cubes.  Regardless, averaging what is something like a power-law distribution can be problematic, of course.

This is reassuring, since in contrast the measured velocity dispersion
is known to critically depend on the adopted percolation length in the
FOF algorithm \citep{nw87}, and often shows an unwanted dependence on
distance.  One reason why our cluster-finding algorithm is more robust
than percolation methods in redshift space is that cluster membership
is always determined within a limited window around the known cluster
center, rather than relative to the last merged unit.  This suppresses
the non-physical structures sometimes identified by the FOF method,
such as apparent long filaments that are often due to two physically
unrelated systems linked in redshift space by member galaxies with
extreme peculiar velocities.

\section{Statistical tests of the algorithm}

Having defined an algorithm, we now test its stability and effectiveness 
for robustly detecting
clusters over a wide range of richness and over a broad redshift
baseline.  Such  a comparison requires great  care; the identification
of clusters in redshift space  and the measurement of their parameters
in  an  unbiased way  are  inherently  difficult,  even in  the  local
universe.  Since  the galaxy correlation  function is positive  over a
wide  range  of  physical  scales, cluster  definition  is  inevitably
sensitive to  the selection window.   
However, in what follows,  we will
show that some  cluster parameters may be less sensitive  to how clusters
are identified  than others.

We are concerned in this section only with testing our method's
potential performance; to this purpose, we use all the galaxies
within each mock catalog that satisfy the magnitude
and redshift conditions $I_{AB} \leq 23.5$ and $0.7 < z < 1.2$.  We
defer discussion of the DEEP2 target selection bias, which limits
the number of galaxies for which the survey will be able to provide
redshift measurements in dense regions, to $\S$ 5.

In figure \ref{fig10} we present a mock catalog for one DEEP2 field
collapsed along the smallest axis, covering the redshift range
$z$=0.7-1.2 and containing a volume of roughly $10^6 h^{-3}$ Mpc
comoving ($\sim$ 15000 galaxies).  We also show the real-space
distribution of members of those FoF groups with $N_{mem}\geq 5$
elements above the survey magnitude limit (central panel) and the
reconstructed population of galaxies assigned to systems with at least
5 elements by our Voronoy-Delaunay algorithm operating in redshift
space (lower panel).  In figure \ref{fig11} we show the 2-dimensional
sky distribution using angular coordinates for the same sample of
objects.  Using these graphic projections we can qualitatively compare
the space distributions of real structures defined above with the
space distribution of clusters recovered in the redshift space
flux-limited sample by the VDM.  Note how in both cases the large
scale patterns defined by galaxy systems and the associations of
clusters in higher-order structures reproduce the underlying real
space landscape.

To assess the performance of our cluster-finding algorithm, we
can test the statistical consistency between the distributions of the
cluster parameters of interest recovered in redshift space and the
corresponding distributions inferred from those members of the
clusters identified in real space that are above the survey magnitude limit.

We use several physical properties as indicators to probe the
reliability and effectiveness of the VDM: viz., richness, velocity
dispersion, mass, and redshift distribution.  For instance, figure
\ref{fig12} shows how well we are able to reproduce the actual
cumulative distribution function for the richness of clusters in one
of our mock catalogs.  The same level of accuracy is achieved in each
mock catalog.  We note that the total number of systems recovered
by the VDM 
%CCT: Above what velocity /N_mem limit? or is this total (as I have assumed)?
when the parameter ${\cal R}\sbr{min}$ is varied up to 20\%
of its assumed value (see \S 3.2.1) is always within 20\%
of the real total number of systems.  This insures that the VDM,
besides the intrinsic scaling, also preserves the absolute
normalization of the relation.

A  quantitative   test  of  the  one-to-one  correspondence  between
reconstructed and real groups can  be performed by counting the number
of members of each reconstructed  cluster that belong to a common real
group. We first  determine which members of each  real group belong to
the  same reconstructed  systems.  We may then  define  the largest  group
fraction (LGF) for  each real group by dividing  the number of members
in its largest redshift-space subgroup  by the total number of members
in the  real group (see  \citet{frea} and \citet{giu00} for  a similar
definition of the LGF).  If  a cluster is identified in redshift space
as a single  system with all of  its members, it would have  an LGF of
100\%, while if none of its  members were assigned to groups in common
its LGF would approach 0.
%\hspace{19cm}
\vbox{%
\begin{center}
\leavevmode
\hbox{%
\epsfxsize=10.4cm
\epsffile{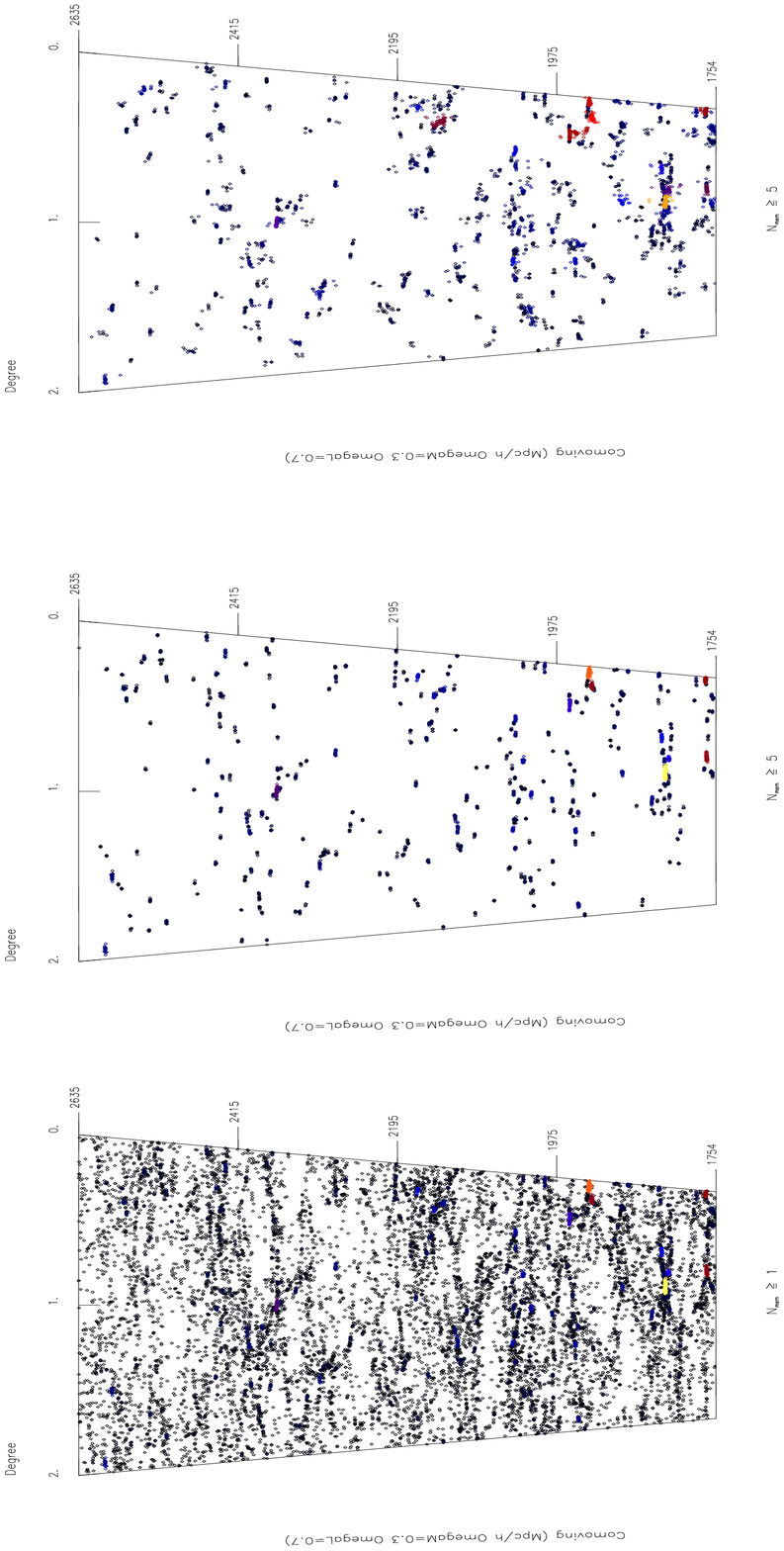}}
\begin{small}
\figcaption{%
{\em Left:}  all galaxies in 
one of our DEEP2 mock catalogs (\#1) represented in a 
2D, real-space cone diagram. The surveyed volume 
corresponds to an angular area of 1 square deg and to 
the redshift interval z=0.7-1.2 (here expressed in \hmpc ~units).
{\em Center:} real space, large-scale spatial distribution of galaxies belonging to 
clusters with more 
than 5 members as identified using the FOF
algorithm. {\em Right:} real space, large-scale spatial distribution of 
galaxies belonging to clusters with more 
than 5 members as reconstructed by the VDM algorithm in redshift space. 
\label{fig10}}
\end{small}
\end{center}}
\vbox{%
\begin{center}
\leavevmode
\hbox{%
\epsfxsize=8.9cm
\epsffile{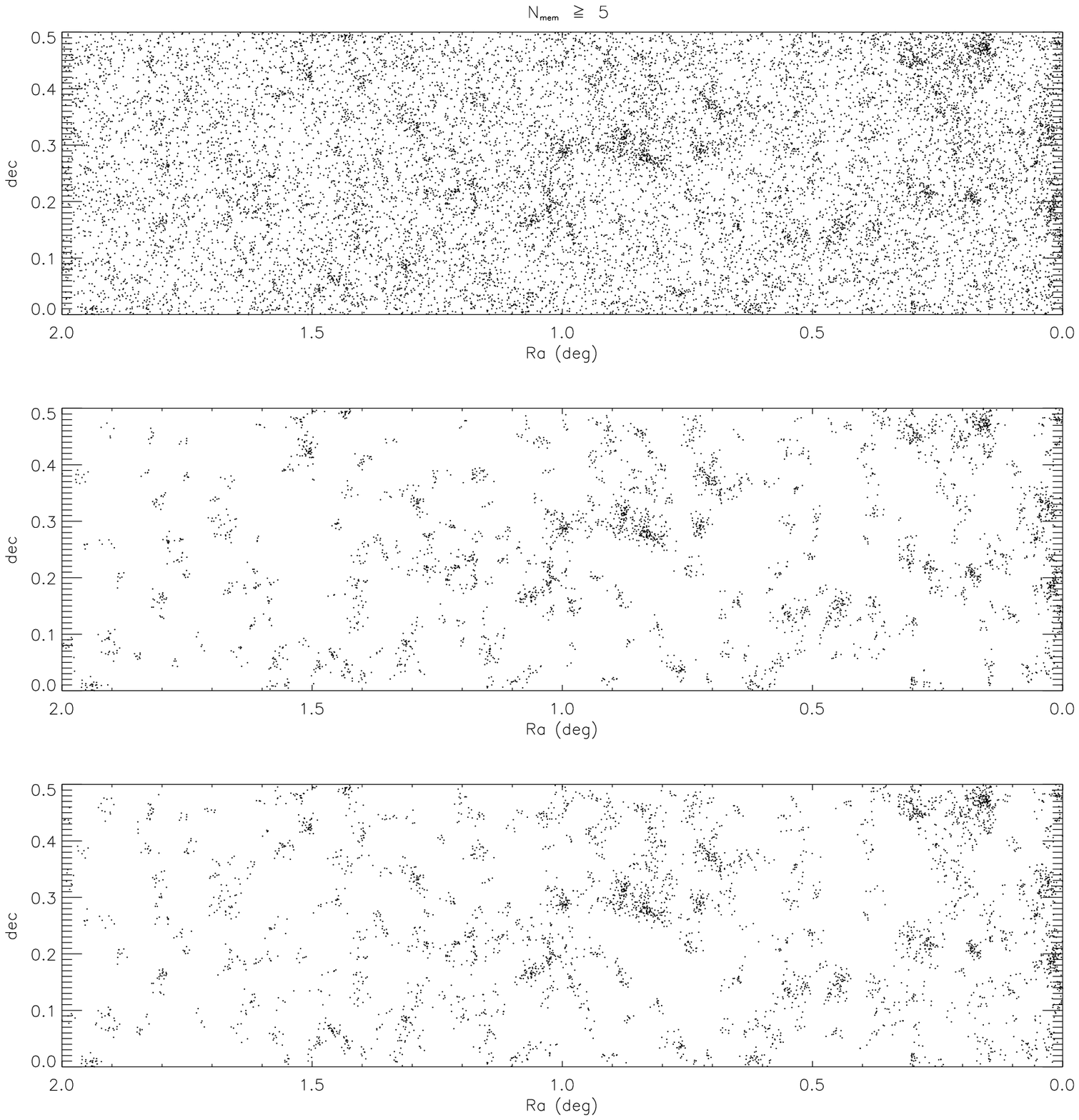}}
\begin{small}
\figcaption{%
{\em Upper:} sky distribution in angular coordinates of all the galaxies
brighter than the DEEP2 limit extracted from  mock  catalog \#1.  
{\em Center:} sky distribution of those galaxies belonging to real-space identified groups with $N_{mem} \geq 5$ elements.
{\em Lower:}  sky distribution of galaxies associated to groups with at least 5 elements as determined by our VDM algorithm applied in redshift space. 
\label{fig11}}
\end{small}
\end{center}}

\vbox{%
\begin{center}
\leavevmode
\hbox{%
\epsfxsize=8.9cm
\epsffile{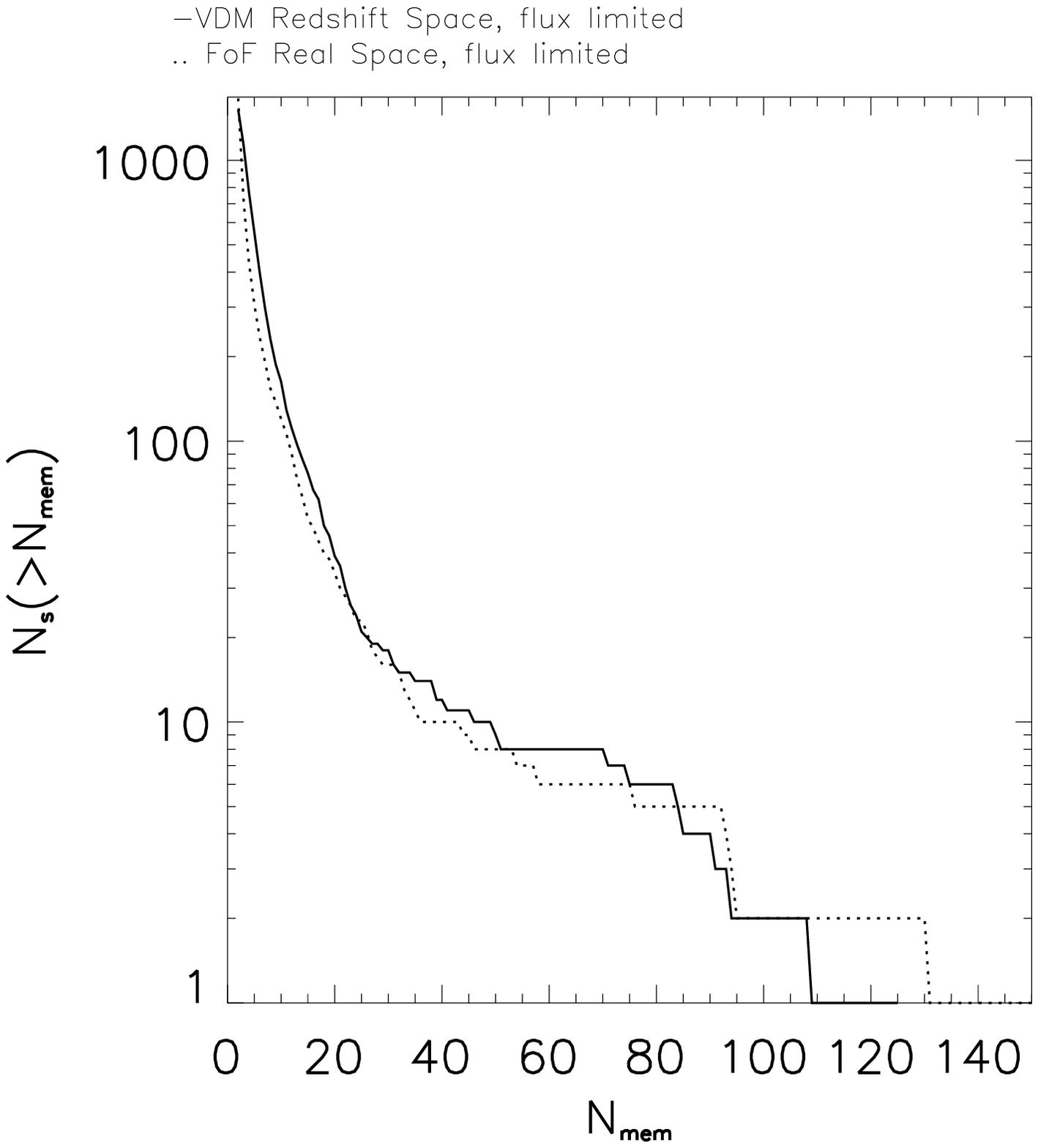}}
\begin{small}
\figcaption{%
%CCT REVISED CAPTION
Cumulative distribution functions (in logarithmic units) of the real
or reconstructed system richness of clusters in mock catalog \#1.
\label{fig12}}
\end{small}
\end{center}}

\vbox{%
\begin{center}
\leavevmode
\hbox{%
\epsfxsize=8.9cm
\epsffile{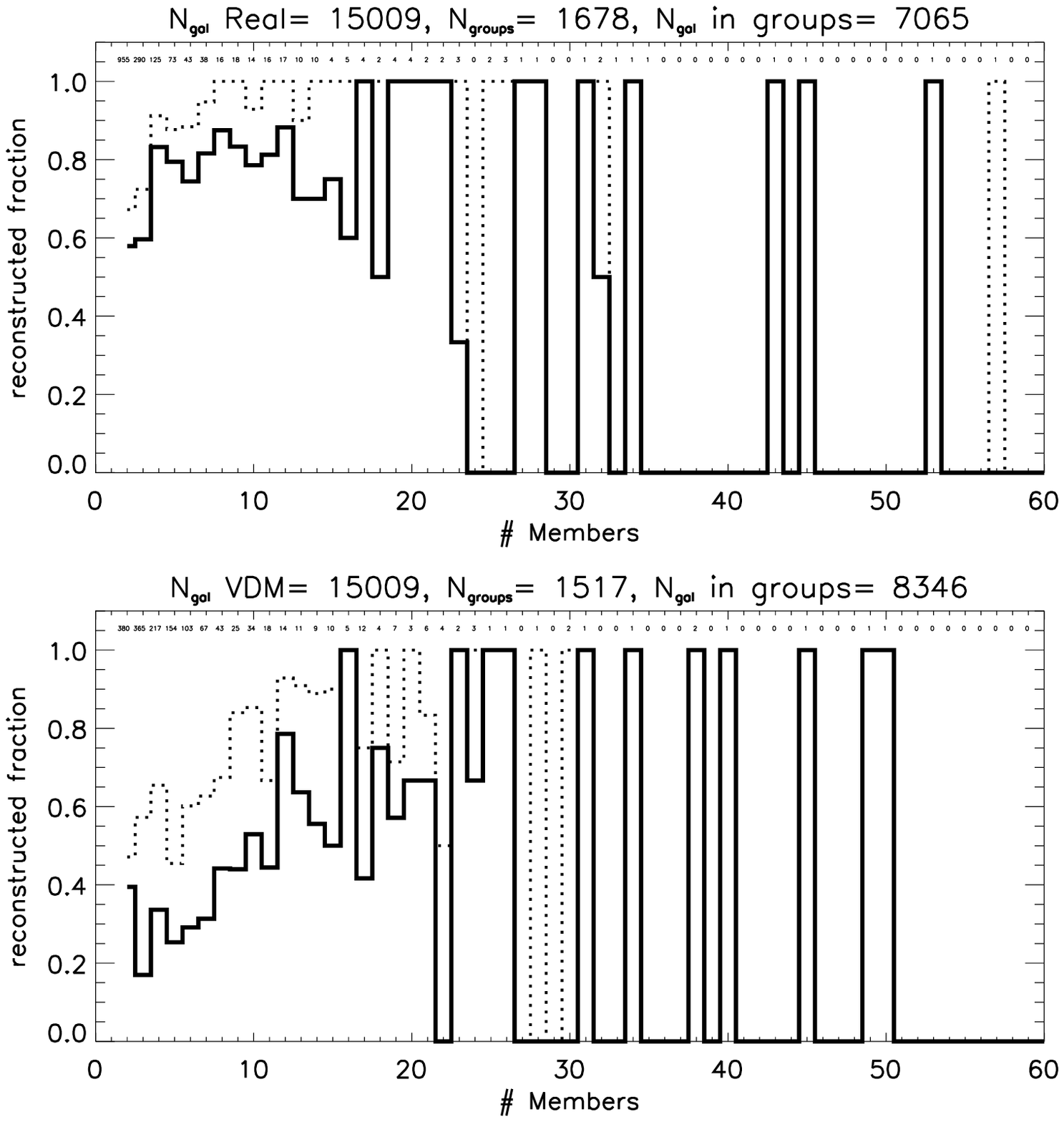}}
\begin{small}
\figcaption{%
{\em Upper:} histogram of the largest group fraction (LGF; see \S 4 for definition) as a function
of the number of galaxy members in the groups identified in real space.  
The solid line gives the
fraction of redshift-space recovered groups with LGFs 
between 75\% and 100\%, while the dotted line corresponds
to LGFs greater than 50\%. The numbers at the top of each 
bar is the total number of systems in real space with that given number 
of members. {\em Lower:} the largest group fraction  as a function
of the number of galaxy members in groups identified in redshift space. 
\label{fig13}}
\end{small}
\end{center}}
  The upper panel of figure \ref{fig13} shows, as a
function of group  richness (number of members), the  fraction of real
groups having a given LGF.  For example, there are 15 real groups with
10  elements; of  these,  70\% have  75\%  - 100\%  of their  elements
identified as  belonging to  the same group  in redshift  space, while
80\% of  them have an  LGF of at  least 50\%. 
The large  fraction of
groups having high  LGF values provides a quantitative  example of the
effectiveness of our cluster membership determination.  Symmetrically,
we can  define the LGF of  systems identified in  redshift space using
their subgroups  found in  real space.  The  histogram of  the largest
group  fraction  as  a  function  of  the  number  of  galaxy  members
identified in redshift  space is shown in the lower  panel of the same
figure. Here we can  appreciate the contribution of interlopers, which
tend to spuriously increase the number of members in the reconstructed
group for systems with less than 30 elements.  However, again the high
fraction of groups  with LGF of at least 50\%  confirms the utility of
our strategy.

A useful cluster sample must satisfy well-defined selection criteria,
possibly maximizing the number of objects meeting the completeness
limits.  In many dynamical studies it is imperative to collect a
sample of clusters complete with respect to the velocity dispersion
parameter \citep{bggg97}.
\vbox{%
\begin{center}
\leavevmode
\hbox{%
\epsfxsize=8.9cm
\epsffile{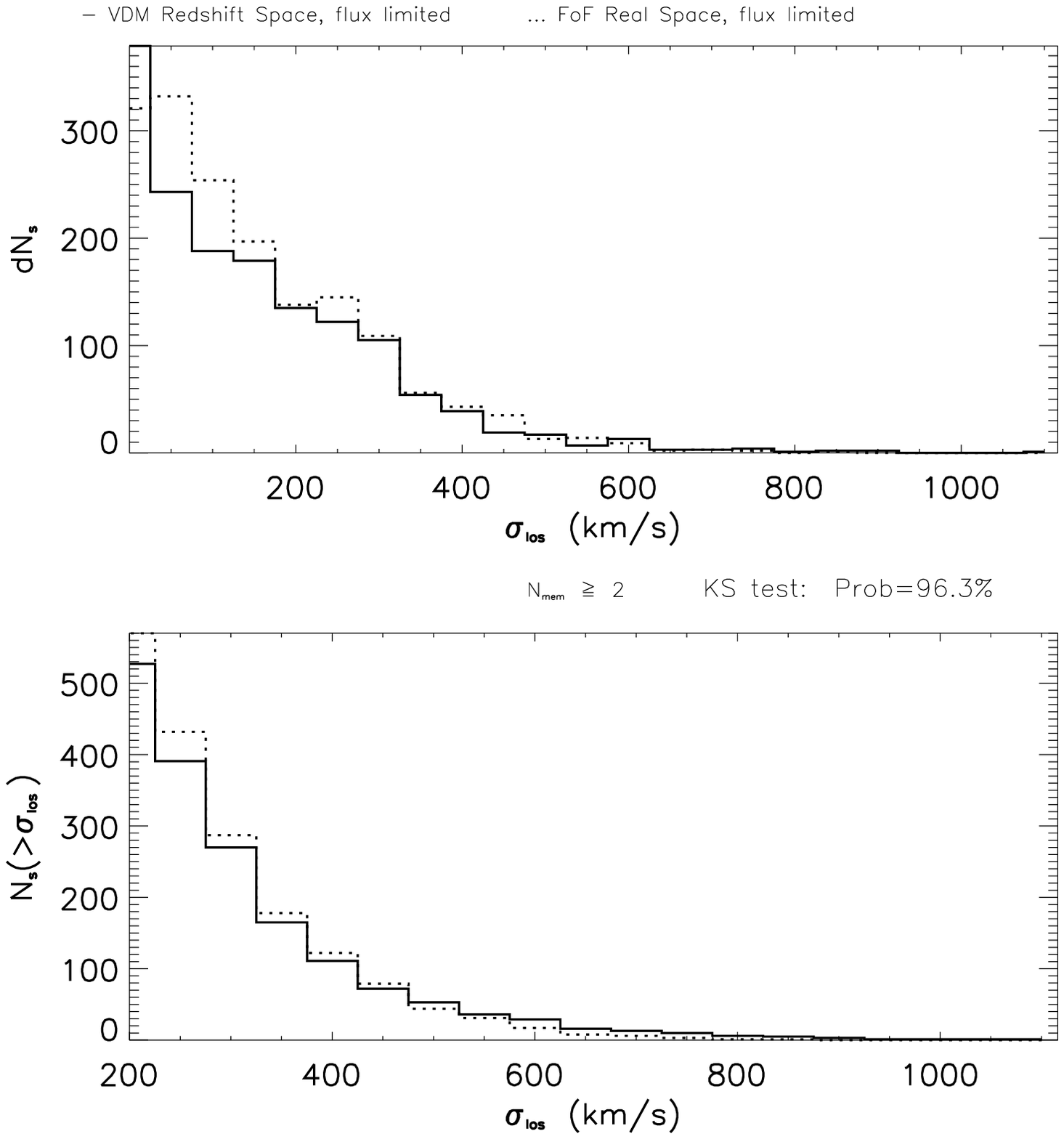}}
\begin{small}
\figcaption{%
Differential ({\em upper}) and cumulative ({\em lower}) distributions of the number
of clusters with $N_{mem}\geq2$ detected and reconstructed  in mock  catalog \#1
as a function of their 1D projected velocity dispersions.
The distribution of clusters identified by FOF in real space is shown by the  dotted line, while the 
distribution recovered by VDM in redshift space is the solid line. 
Data are binned  in velocity dispersion  intervals of width 50 \kms.  
A Kolmogorov-Smirnov test confirms the consistency.  
\label{fig14}}
\end{small}
\end{center}}
However, most workers have selected samples
complete to some richness, and assumed that this translates into a
completeness in $\sigma_{los}$, although it is well known that samples
collected in this way are usually biased towards low $\sigma_{los}$
values \citep{maz96, bi98}.  As we are most interested in the cluster
velocity function, we adopt a different approach here.
 
In figure \ref{fig14} we plot both the differential and cumulative
distribution of the number of clusters found in one mock DEEP2 field
as a function of the projected line of sight velocity dispersion
$\sigma_{los}$ for both the real-space FoF and redshift-space VDM
samples. A Kolmogorov-Smirnov statistical analysis (KS) quantifies the
level of agreement between the intrinsic cumulative velocity function
of the DEEP2 sample and the one recovered in redshift space. The plot
shows how the VDM can reliably identify an homogeneous set of systems
which reproduce real clusters characteristics down to $\sigma_{los}
\approx 200$ \kms. These results must be compared to standard
techniques of cluster identification and interloper rejection which
are systematically affected by observational biases for $\sigma_{los}
\leq 600$ \kms \citep{bggg97}.  

 By analyzing the six different mock
catalogs, we find that the average number of real (FoF) clusters with
$N_{mem}\geq 2$ members a simulated DEEP2 field is $601 \pm 43$, while
the average number of systems reconstructed in redshift space by the
VDM is $571 \pm 46$, with a satisfactory average KS probability value
of $ (55 \pm 35)\%$. We may check the stability of the final cluster sample properties 
by  varying  the clustering parameter ${\cal R}\sbr{min}$. 
\vbox{%
\begin{center}
\leavevmode
\hbox{%
\epsfxsize=8.9cm
\epsffile{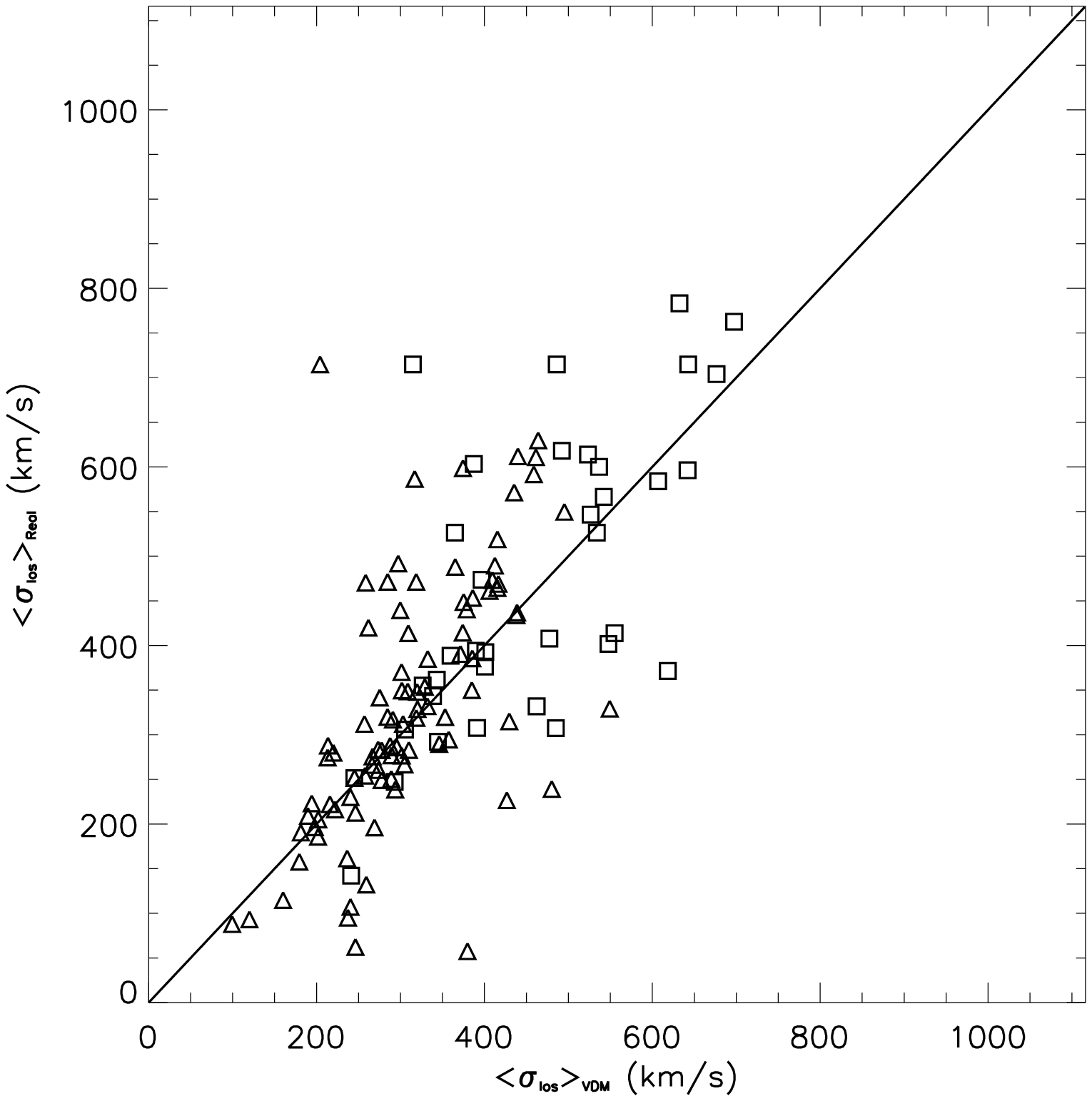}}
\begin{small}
\figcaption{%
Real and recovered velocity dispersions of systems with $N\sbr{mem}
\geq 10$ elements are compared.  For each recovered system, the
corresponding real-space structure is identified by determining which
group members belong to the same structure in real space and using as
a reference the largest of these real systems. Systems with 10$\leq N
\leq 20$ and with $N >20$ members are indicated by triangles and
squares, respectively.
\label{fig15}}
\end{small}
\end{center}}

If  we lower/increase the optimal value ${\cal  R}\sbr{min}=1$ \hmpc ~by 20\%
the  cumulative  and  differential  distribution functions  are  still
consistent with  being unbiased  down to 200  \kms ~(the  KS associated
probability   drops  to  $(40 \pm 33)\%$/$(50\pm 30)\%$   respectively),
but the overall number of reconstructed systems would also 
decrease/increase by nearly 20\%, for a total of 
$462 \pm 20$ ($700 \pm 50$) systems per square degree.

The overall performance of the method can also be seen in figure
\ref{fig15} where we plot the intrinsic and recovered velocity
dispersions on a cluster-by-cluster basis. Here we can see that
occasionally a large system is fragmented into multiple components,
some with very few members and low velocity dispersions. In an
opposite sense, interlopers tend to increase the estimated velocity
dispersion of smaller systems.  This bias, which mostly affects the
percolation algorithms, acts to steepen the slope of the
$\sigma-\sigma$ relation.  However our algorithm, by locally defining the clustering parameters, is able to
counteract this tendency and elastically perform over a wide range of
velocity dispersions.  While the
characteristics of the systems reconstructed using the FOF method are
known to be fairly sensitive to the scale length of the adopted
linking parameters (Frederic 1995a,b, Giuricin et al. 2000), there is
no preferred velocity or richness scale artificially introduced into
the VDM reconstructed systems; as a consequence the system velocity
dispersion can be efficiently used as a stable and unbiased parameter
to define a complete cluster sample (see \S 6).

\vbox{%
\begin{center}
\leavevmode
\hbox{%
\epsfxsize=8.9cm
\epsffile{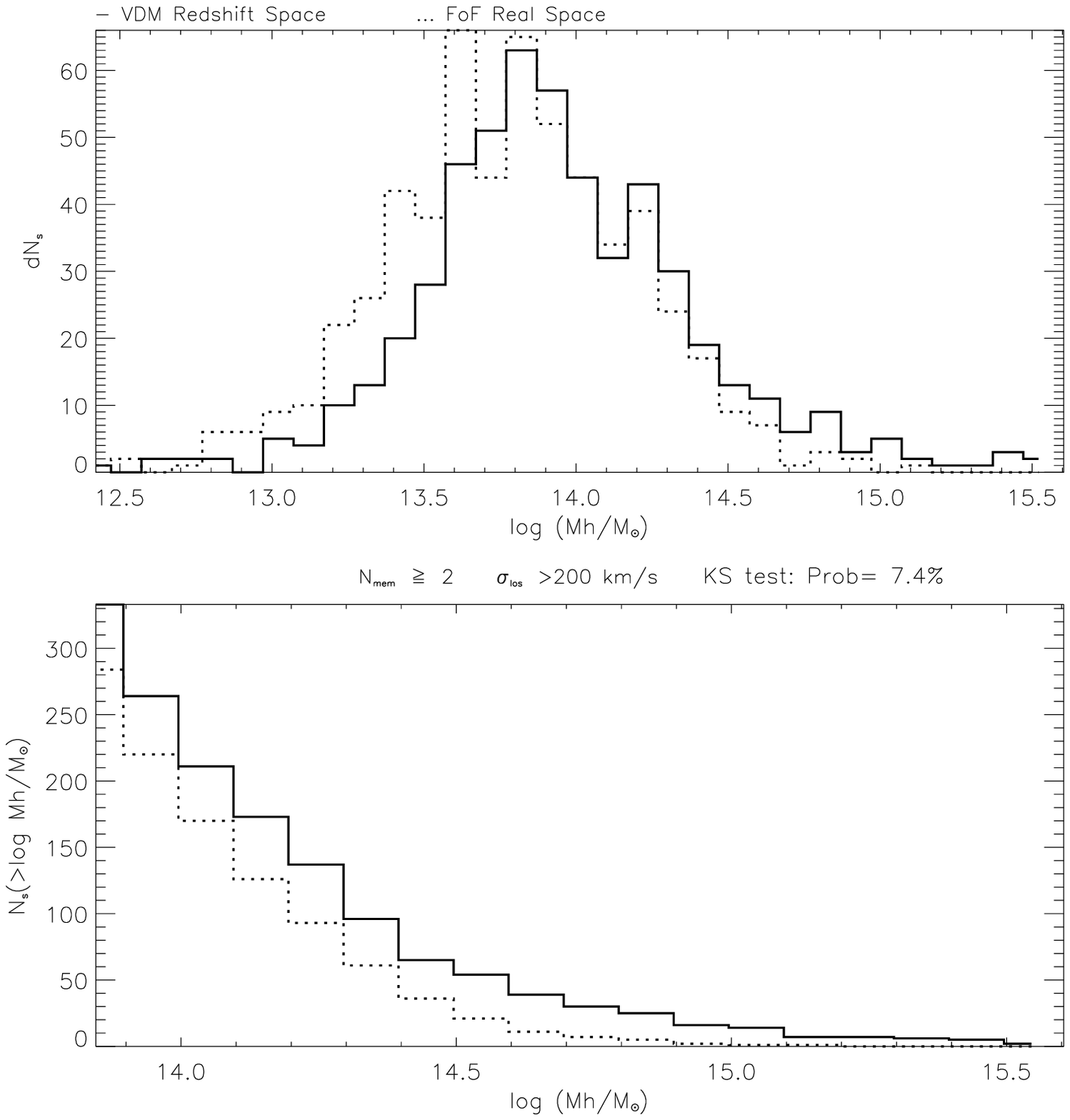}}
\begin{small}
\figcaption{%
Differential ({\em upper}) and cumulative ({\em lower}) distributions of the number
of clusters detected and reconstructed in mock catalog \#1
as a function of their virial mass.
The mass is computed using to the median mass estimator 
of \citet{htb85}. The  distribution of real-space groups is shown by the dotted line,
while the observed 
distribution recovered in redshift space is the solid line. Data are
binned in logarithmic mass intervals of width 0.1. 
\label{fig16}}
\end{small}
\end{center}}

We next compute the virial mass of a system recovered in redshift
space by the VDM and compare with the intrinsic value (see Appendix
A).  We are not concerned here with any biasing scheme that may relate
the mass inferred from optical tracers to the true halo mass; we only
investigate to what extent the virial analysis applied in the presence
of a distance degeneracy along the line of sight is able to reproduce
the virial mass estimate from the members of the same group identified
in real space that are above the DEEP2 flux limit.

The hypotheses and approximations at the heart of the virial analysis
are known to oversimplify reality \citep{bt81}.  Moreover the virial
mass critically depends on the estimated cluster harmonic radius (see
Appendix A), and it is easy to show that potential interlopers are the
dominant source of systematic errors.  While the galaxy distribution
inside a cluster is described by a radially decreasing function
(generally assumed to be a $r^{-2}$ profile) the probability of
interlopers increases as $r^2$, leading to a systematic tendency to
overestimate the harmonic radius of the recovered cluster.  Even with
a quadratic dependence on a reliable estimator such as the velocity
dispersion, the mass estimate will be linearly affected by this
systematic offset.  This effect can be appreciated in figure
\ref{fig16} where we plot the differential and cumulative mass
distributions for one of our mock simulations.  

Even when the mass is evaluated only for in the regime where the
velocity function is recovered accurately ($\sigma_{los} \gtrsim 200$
\kms), we observe a systematic offset between the real- and
redshift-space samples.
\vbox{%
\begin{center}
\leavevmode
\hbox{%
\epsfxsize=8.7cm
\epsffile{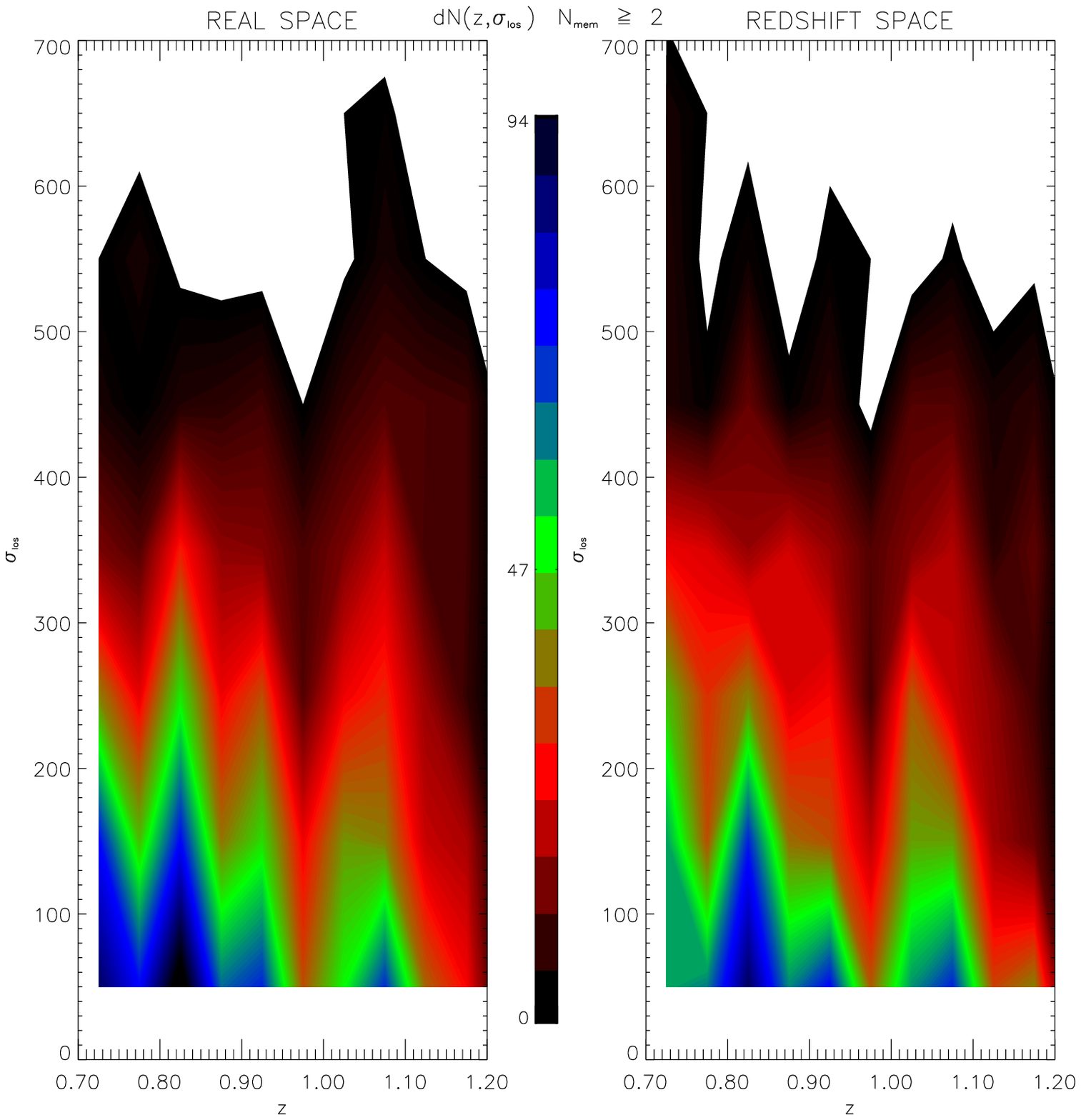}}
\begin{small}
\figcaption{%
Differential number counts of clusters as a bivariate  function of their redshift z and 
projected velocity dispersion $\sigma_{los}$. The parent distribution 
of cluster counts identified in real space is shown in the left panel, while 
the observed distribution recovered in redshift space with our method 
is plotted in the right panel. Density levels, smoothed by 0.05 in z and 50 \kms ~in 
$\sigma_{los}$, are represented by means of  the colorbar.
\label{fig17}}
\end{small}
\end{center}}

This effect is only marginally corrected for
by applying more robust methods such as the median mass estimator of
\citet{htb85} (see eq. 12 in that work).  A better agreement between
real and reconstructed mass distributions may be achieved by lowering
${\cal R}\sbr{min}$ by 10\% from its nominal value (1 \hmpc).  In this
way a satisfactory degree of agreement is reached for systems with $M
\geq 10^{14} h^{-1}M_{\odot}$. In analyzing all the six mock catalogs
we then reconstruct an average of $160\pm 22$ groups per square degree
out of the original $193\pm8$ real systems with accurate mass
statistics.

In a previous paper \citep{new01} we showed that the evolution of the
comoving abundance of clusters as a function of their velocity
dispersion $\sigma$ and redshift $z$ is a sensitive function of
cosmological parameters.  The statistical significance and robustness
of the cluster abundance test depends critically on an unbiased
mapping of the distributions of the cluster observables from real to
redshift space.  How well this can be achieved is seen in figure
\ref{fig17} where we show that the reconstructed statistic is a
satisfactorily undistorted reproduction of the underlying real
$N(\sigma,z)$ function.  The large number of recovered systems,
together with the small fluctuations in the total number of clusters
observed in the six independent mock catalogs (reflected by their
small variance, $\sim$7\% of the mean value) should allow us to
perform the test without systematic biases.  It is necessary only to
tune ${\cal R}\sbr{min}$ with a reasonable set of simulations to
ensure that both the total number of clusters and their distribution
will be reconstructed successfully over some range in velocity dispersion.

\vbox{%
\begin{center}
\leavevmode
\hbox{%
\epsfxsize=8.9cm
\epsffile{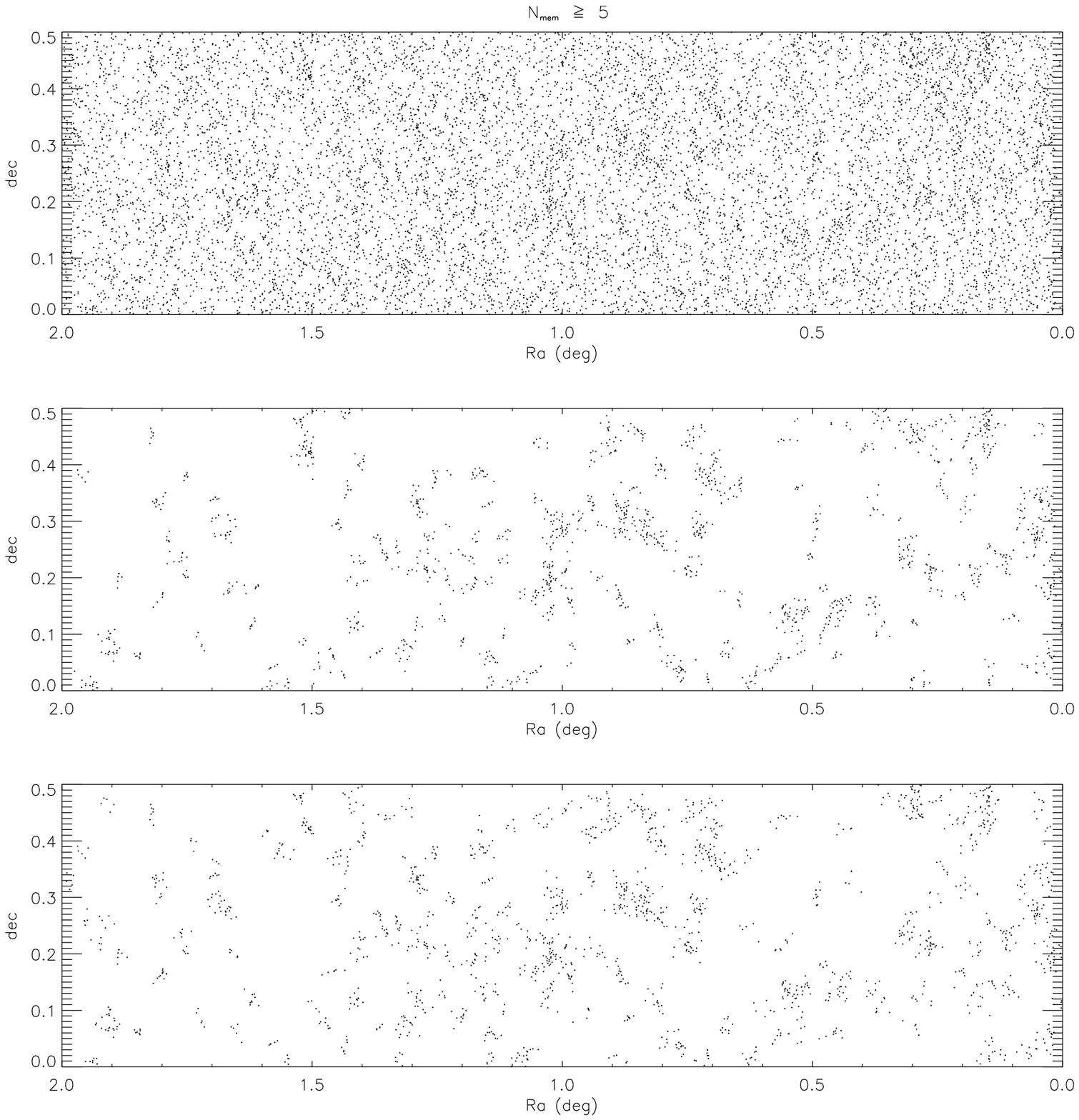}}
\begin{small}
\figcaption{%
{\em Upper:} sky distribution of DEEP2 target galaxies selected to be
on a slitmask for mock catalog \#1, also represented in fig
\ref{fig11}.  {\em Center:} sky distribution of mock catalog \#1
selected galaxies belonging to real-space groups with more than 5
elements assigned to DEEP2 masks.  {\em Lower:} sky distribution of
galaxies associated to groups with more than 5 elements by our VDM
algorithm applied to mock catalog \#1 selected targets only.
\label{fig18}}
\end{small}
\end{center}}

\section{Effects of DEEP2 target selection bias}

In the four DEEP2 fields, individual slitmasks will each cover
overlapping 16$\arcmin$ by 4$\arcmin$ regions.  There will be
$\sim$130 slitlets per mask, but the mean surface density of candidate
galaxies still exceeds the number of objects we can select, and
spectra of selected targets cannot be allowed to overlap on the CCD.
Spectra will thus be obtained for $\sim$70$\%$ of the galaxies in each
field that meet our color and magnitude selection criteria, with the
specific objects chosen by a slitmask algorithm which is necessarily
biased against the highest-density regions, where the spectra of
neighboring galaxies would overlap on the CCD.  Here we test how this
bias, which forces us to observe a lower fraction of targets in the
densest regions (like the cores of clusters) will affect VDM results.
We can use the same sample to investigate to what velocity dispersion
the sample of clusters recovered from DEEP2 will be complete.

Figure \ref{fig18} shows the results of our target selection on 
mock galaxy  catalog \#1.  The upper  panel shows the sky
distribution of galaxies  which would be targeted to  be on a slitmask
(mock \#1  targets).  A comparison of this galaxy  map with the upper
panel of figure \ref{fig11}, which shows all the galaxies 
in that field, shows  the effects of our target selection
strategy which  is necessarily most strongly  biased on small projected  scales. Note the smoother spatial density contrast  compared to the clumpy 
distribution in figure \ref{fig11}.

The specific DEEP2 observing strategy can be easily incorporated when
applying the VDM; we simply increase the value of ${\cal R}\sbr{min}$
by a factor $s^{-1/3}$, where $s$ is the fraction of all target
objects that are selected for observation. 
\vbox{%
\begin{center}
\leavevmode
\hbox{%
\epsfxsize=8.9cm
\epsffile{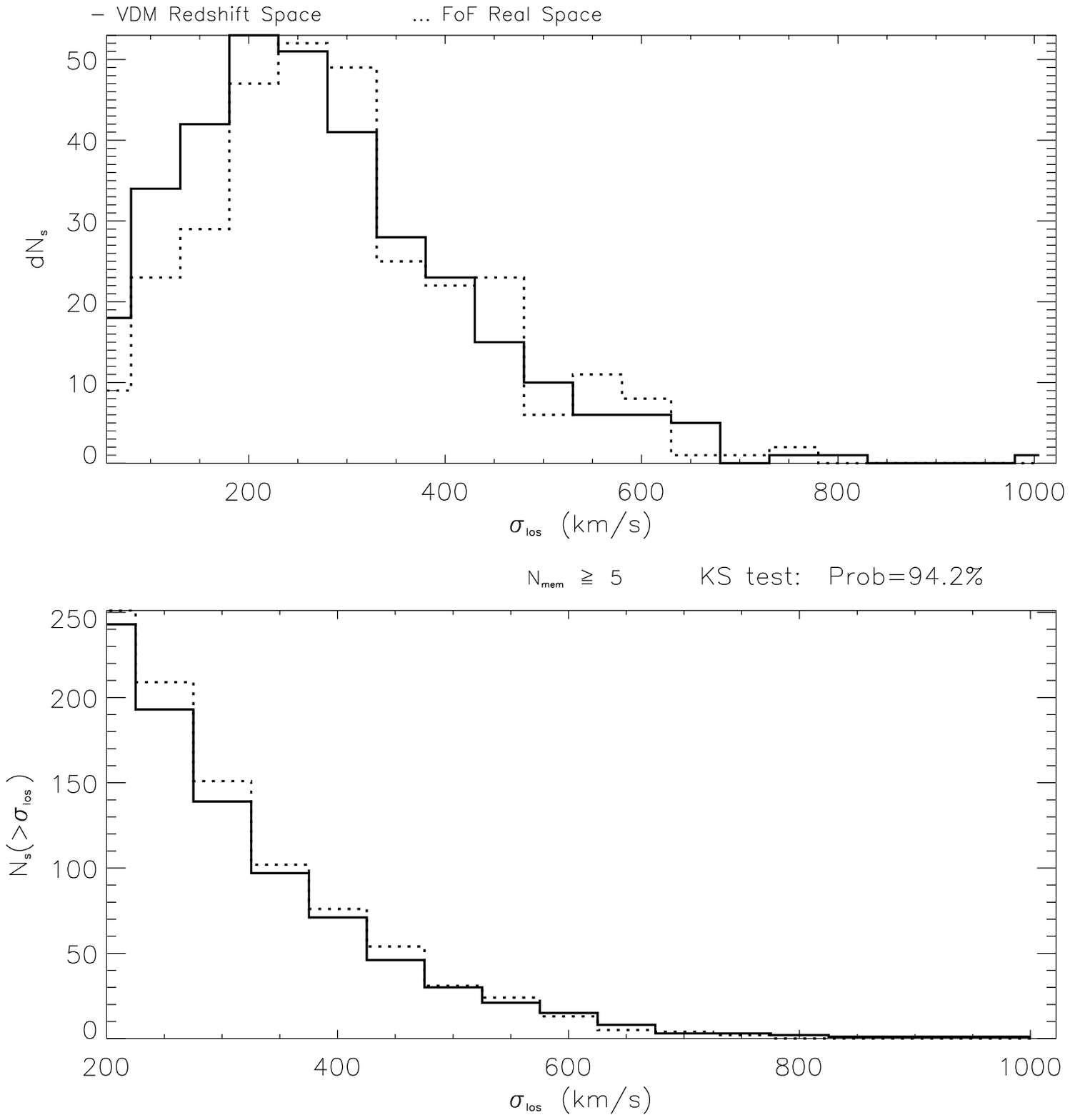}}
\begin{small}
\figcaption{%
Differential ({\em upper}) and cumulative ({\em lower}) distributions
of the number of clusters with $N_{mem}\geq5$ elements found in mock
catalog \#1 as a function of their projected velocity dispersions.
The distribution of velocity dispersions derived from all galaxies
meeting the DEEP2 magnitude limit in each system in the real-space
mock catalog is shown by the dotted line, while the observed
distribution recovered by VDM in redshift space after applying the DEEP2 slitmask target selection criteria is plotted as a solid line.
Data are binned in velocity dispersion intervals of width 50 \kms.  A
Kolmogorov-Smirnov test confirms the consistency.
\label{fig19}}
\end{small}
\end{center}}
This counteracts the fact
that our sampling of necessity dilutes the central regions of clusters
and thus could bias the algorithms in the detection phase (\S 3.2.1).
In the central panel of figure \ref{fig18} we plot the map of galaxies
assigned slitlets by our selection procedure and which belong to
real-space groups with at least 5 members selected for observation,
while in the lower panel we plot the galaxies belonging to groups
($N_{mem} \geq 5$) identified by applying the VDM to the
redshift-space catalog of selected targets in mock catalog \#1.  These
plots graphically show that the map of targeted cluster members
detected by the algorithm traces the same structure as the map of
galaxies which are known to be members of real clusters in the
targeted sample.

We next investigate how the DEEP2 target selection biases affect the
$\sigma_{los}$ threshold above which cluster members are reliably
identified and reconstructed by the VDM.  In figure \ref{fig19} we plot the
distribution of the number of clusters as a function of their line of
%CCT: make sure the description of the real-space sample below is right.
sight velocity dispersion.  Both the real distribution (derived
including all galaxies meeting the DEEP2 magnitude limit, not just
those assigned slitlets) and the distribution recovered by the VDM
after applying to the sample our slitmask design algorithm (mock catalog \#1
targets) are shown.  The threshold of unbiased identification,
$\sigma_{los} \approx 200$ \kms, is now reached only if we 
implement a second selection parameter which discriminates groups
according to their richness. 
\vbox{%
\begin{center}
\leavevmode
\hbox{%
\epsfxsize=8.9cm
\epsffile{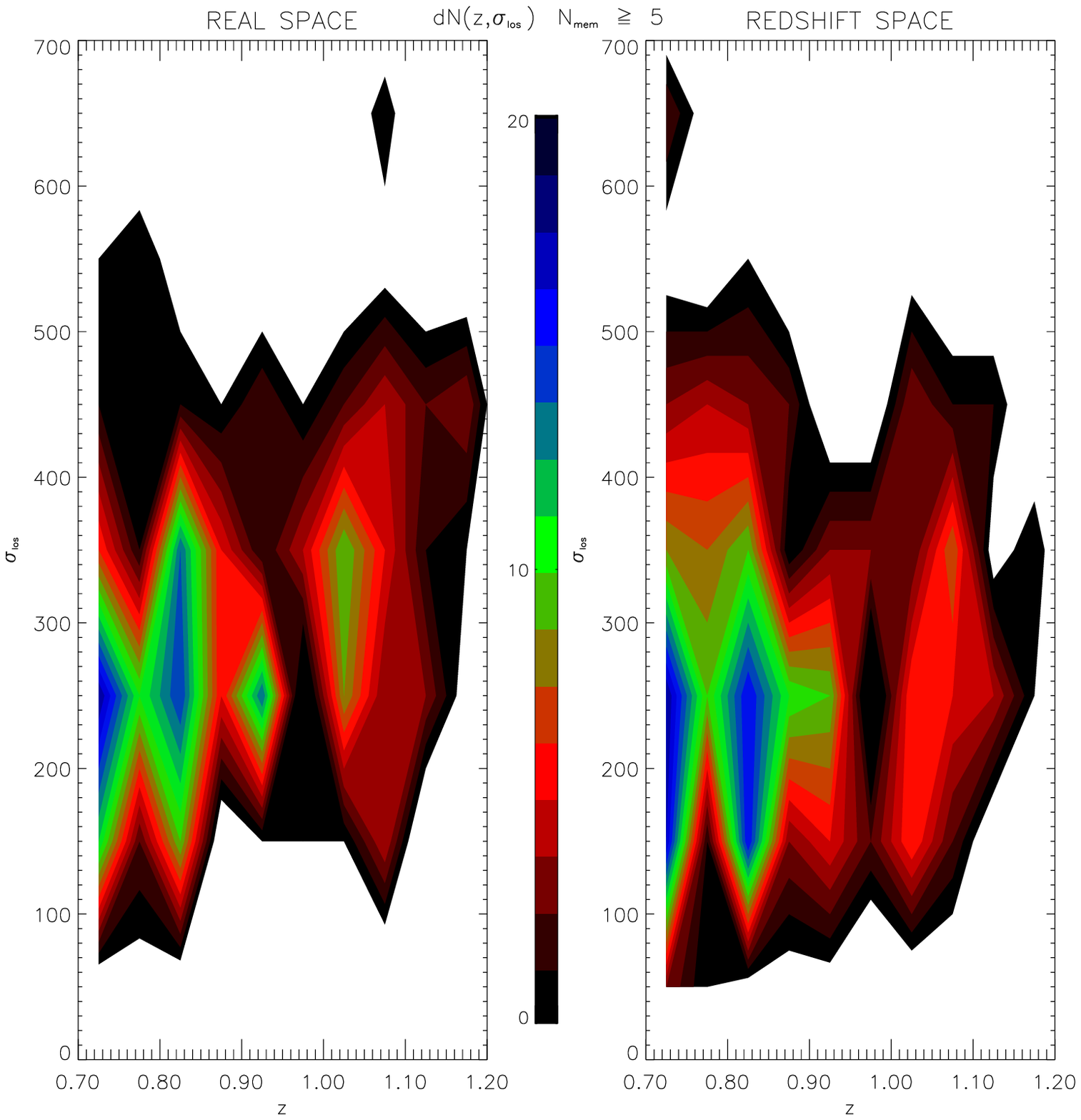}}
\begin{small}
\figcaption{%
Differential number counts of clusters as a simultaneous  function of their 
redshift z and 
projected velocity dispersion $\sigma_{los}$ for systems with $N\sbr{mem} \geq 5$
members.
The parent distribution  of cluster counts identified in real space is shown 
on the left, while 
the observed distribution recovered by VDM in redshift space after applying 
to the galaxy sample the DEEP2 target selection criteria
is plotted on the right. Density levels, smoothed by 0.05 in z and 50 km/s in 
$\sigma_{los}$, are represented by means of  the colorbar.
\label{fig20}}
\end{small}
\end{center}}
An acceptable fit between real and reconstructed
distributions (average KS probability of ($35 \pm 30)$\% for the six
mocks) is obtained if we require $N\sbr{mem} \geq 5$.  The average
number of real clusters with $N\sbr{mem} \geq 5$ and $\sigma \geq 200$
\kms ~in the six mocks is $268 \pm 16$ per square degree, while the
average number of systems with the same characteristics
reconstructed in redshift space after applying the DEEP2 target
selection algorithm to the mocks is $271 \pm 13$. By
lowering/increasing the initial ${\cal R}\sbr{min}$ parameter by 20\%
we would have recovered $205 \pm 14$ /$ 340 \pm 11$ systems per square
degree with an average KS probability of ($20\pm24$)\% /($30 \pm
34$)\%.

If we do not place conditions on richness, then the
reconstructed sample matches the real cluster distribution for
$\sigma_{los} \gtrsim 400$ \kms ~(average KS probability $(30 \pm
20)$\%).  The average number of systems meeting this condition in our
six mocks is $135\pm 11$ per square degree while the reconstructed
number is $109\pm 11$.  This selection  would  degrade
the sample statistics by a factor of $\sim 2$.  The actual target selection
used in the DEEP2 survey will incorporate adaptive tiling of the
slitmasks; this has not been included in the discussion here and it
should somewhat reduce selection biases.

We have shown that the VDM algorithm works efficiently without
applying any predetermined selection correction as a function of
redshift. This can be seen in figure \ref{fig20} where we plot the
simultaneous distribution of clusters as a function of both line of
sight velocity dispersion and redshift.  Median values for global
parameters (see appendix A for their definition) characterizing real
and reconstructed groups are listed in table \ref{tab1} for two
specific samples (mock catalog \#1 and mock catalog \#3). We note the
typical offset that characterizes quantities whose definition requires
computation of the system radius. It is also interesting to note that
a quantity seriously biased by the DEEP2 mask design criteria is the
system mass-to-light ratio. This is a joint conspiracy of the bias in
the calculated mass and of the fact that the total luminosity of a
system is severely underestimated when only targets selected for DEEP2
observation are considered.

\section{Completeness with respect to the matter distribution}

From a theoretical perspective it is easier to predict how mass
evolves into structures than to describe how it is converted into
light \citep{koc01}.  Observationally, however, catalogs of groups and
clusters of galaxies represent only visible tracers of the general
underlying matter fluctuations.  If we want to use the abundance of
galaxy clusters to put constraints on cosmological quantities as
suggested by many authors \citep{fbc97, p98, h00, b01, mmm01, new01}
it is therefore imperative to assess the completeness of the
reconstructed group sample with respect to the underlying parent
distribution of matter halos \citep{wk01}.

In section  4  we argued that the 
distribution of clusters as a function of their line of sight
velocity dispersion can be reliably determined using 
our VDM algorithm down to 
$\sigma_{los} \gtrsim 200$ \kms. 
We here investigate to what extent the $N(\sigma_{los})$ distribution 
inferred using a flux-limited sample of galaxies as a tracer of the clusters
reproduces the  statistics of dark matter fluctuations.
The region of  velocity space where the agreement is satisfactory 
determines the range of completeness of our cluster sample and the 
region of  feasibility of cosmological tests based upon it.

Since the VDM algorithm is based upon the use 
of locally determined parameters, it is highly independent of the 
particular cosmological simulation used to test it.
However we expect that the halo occupation number, i.e. the rate at which 
galaxies form in a halo as a function of mass, will vary in different 
cosmological scenarios \citep{ps00,mh01}; as a consequence the 
interval of consistency between halo and cluster
velocity dispersions can be a function of the assumed cosmological model.
We therefore  frame the conclusions of this section within the standard
picture of the $\Lambda$CDM model described in \S 2. Moreover, due to the 
uncertainties affecting prescriptions for galaxy formation, conclusions
we draw in this section may be sensitive to  simulation 
details.

In figure \ref{fig21} we show the distribution of systems in mock
catalog \#1 as a function of the line-of-sight velocity dispersion
parameter inferred from {\em all} the galaxies in the halo, applying
no magnitude limit.  We may expect that these galaxies should be less
biased compared to the dark matter than the $\sim L_{*}$ galaxies that
DEEP2 will observe, and thus should approximate the velocity
dispersion of the mass in each cluster.  Superimposed, we also plot
the distribution of clusters recovered by VDM in redshift space
inferred from the galaxies brighter than the DEEP2 magnitude limit.
Note that in both cases we derive the velocity dispersion using the
same computational scheme; only the tracers are different. 
\vbox{%
\begin{center}
\leavevmode
\hbox{%
\epsfxsize=8.9cm
\epsffile{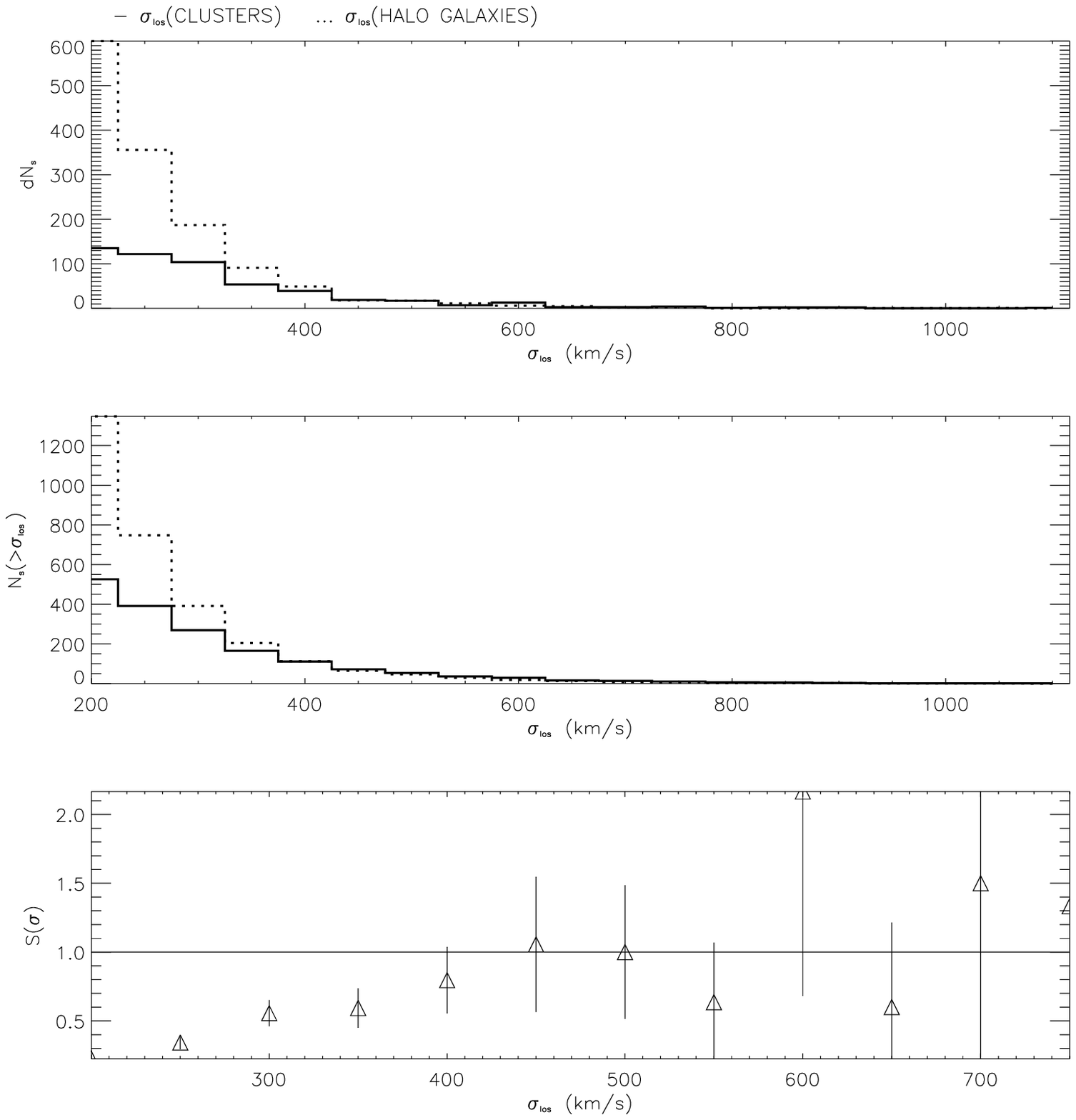}}
\begin{small}
\figcaption{%
The solid lines show the differential ({\em upper}) and cumulative
({\em center}) $\sigma_{los}$ distributions of VDM clusters in mock
catalog \#1 (solid line) is plotted as a function of the projected
velocity dispersion. Also plotted are the $\sigma_{los}$ distributions
of halos identified in real space with a velocity dispersion inferred
using all the galaxies that form within the halo, not just those
meeting the DEEP2 magnitude limit (dotted line).  Data are binned in
velocity dispersion intervals of width 50 \kms.  {\em Lower:} The
sample selection function in velocity dispersion space.  In this
$\Lambda$CDM simulation, clusters recovered by VDM in redshift space
are representative of the general halo population for $\sigma_{los}
\gtrsim 400$ \kms.
\label{fig21}}
\end{small}
\end{center}}
We also
define a velocity dispersion selection function as the fraction of the
clusters identified in redshift space with velocity dispersion
$\sigma_{los}$ out of the total number of matter halos having that
same velocity dispersion. It is clear from figure \ref{fig21} that the
lower threshold of completeness is $\sim 400$ \kms ~below which the
selection function significantly departs from unity.  Above this
limit, the velocity dispersion inferred from the most luminous
galaxies identified as cluster members by our method is consistent
with the parent $\sigma$ distribution of halos, i.e. the effects of
any possible luminosity-dependent velocity bias are minimal.  Note
that to the same velocity limit, the VDM sample is also not affected
by the DEEP2 mask design biases.  It may be possible to use the
derived selection function to correct for unseen halos much in the
same way we correct for unseen galaxies when we compute the galaxy
luminosity function in a flux-limited survey.

The mass distribution of halos can be economically described in terms of
analytical formulas \citep{ps74,st99}.  A similar halo description in
terms of velocity dispersions can be derived assuming a conversion
prescription between the mass M and the one dimensional velocity
$\sigma_{halo}$ \citep{nw88,koc95,new01}. However, the velocity
dispersion of a halo can be computed according to at least two
different operative definitions. 
\vbox{%
\begin{center}
\leavevmode
\hbox{%
\epsfxsize=8.9cm
\epsffile{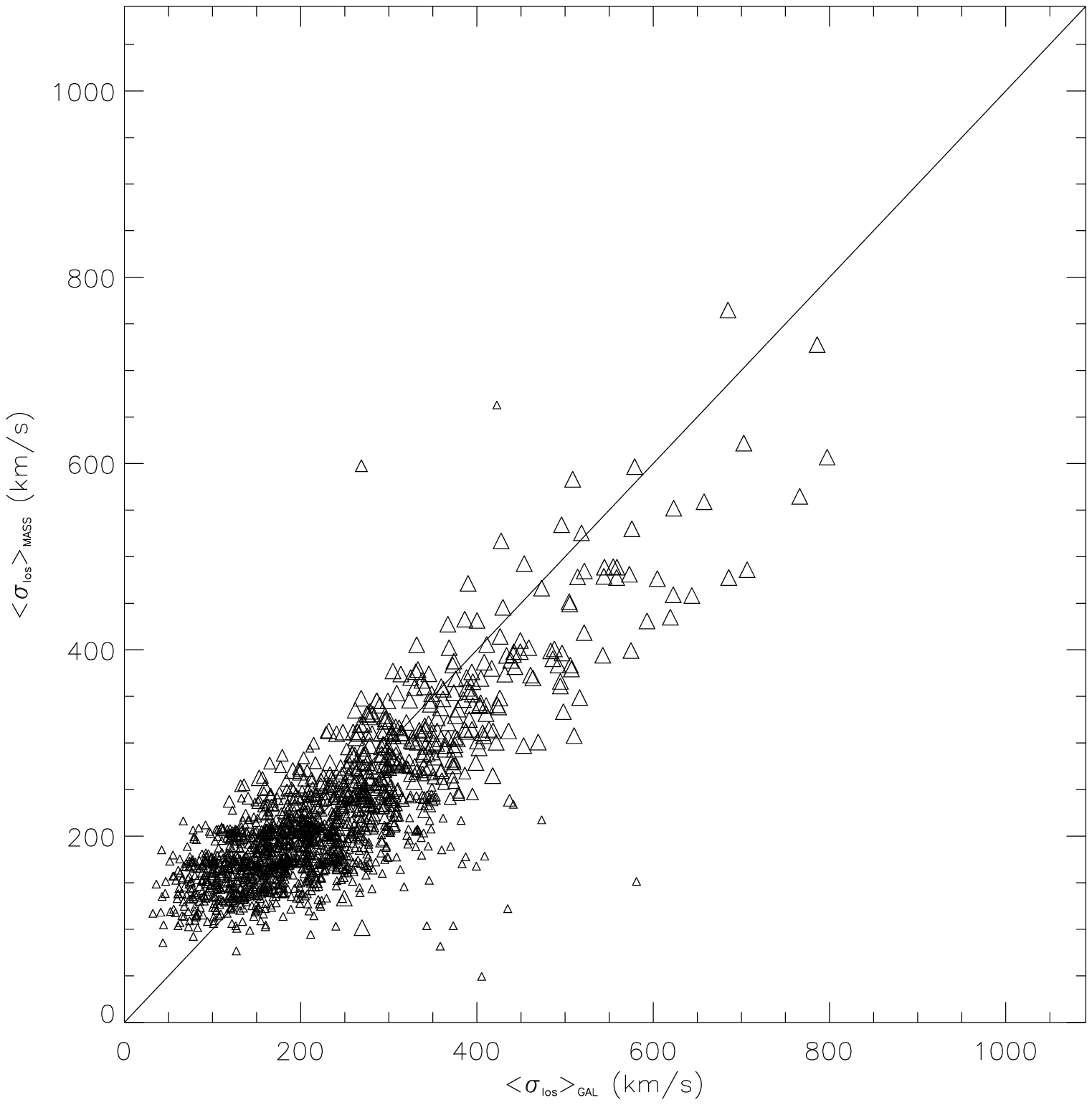}}
\begin{small}
\figcaption{%
Correlation between two estimators of the velocity dispersions of
systems simulated in a $\Lambda$CDM model (for mock catalog \#1; see
\S 2).  Along the x-axis we plot the line of sight velocity dispersion
of the halos inferred applying the estimator described in appendix A
(eq. A3) to all the galaxies that form in the halo
($\sigma_{gal}$). The velocity dispersion inferred using a spherical
isothermal model to describe the halo mass distribution,
$\sigma_{mass}$, is plotted along the ordinate axis.  The triangles
are scaled accordingly to the following three halo richness ranges:
$10 \leq N_{gal} \leq 10$, $10 < N_{gal} \leq 20$, and $N_{gal} > 20$.
The panel shows that the different $\sigma$ estimates are correlated
but offset (the solid line is the curve $\sigma_{gal}$=$\sigma_{mass}$).
\label{fig22}}
\end{small}
\end{center}}
 We can estimate $\sigma_{los}$ by
applying the standard estimator given in Appendix A (eqn. A3) to all
the galaxies that form within a halo.  Alternatively, in the spherical
top-hat collapse model, a halo's mass may be defined in terms of
$r_{200}$, the radius of a spherical volume within which the mean
density is 200 times the critical density at that redshift, and given
by

\[M_{200}=\frac{4 \pi}{3} <\rho(r)>_{r_{200}} r_{200}^3.\]

\noindent In the case of a singular isothermal spherical particle  distribution, 
the density is related to the 
velocity dispersion \citep{bt87} as

\[\rho(r)=\frac{\sigma_{los}^2}{2\pi G r^2};\]

\noindent therefore the one-dimensional velocity dispersion of each
halo can be derived as a function of $M_{200}$ and $r_{200}$. 

In comparing the results of simulations such as those used for this
paper to semi-analytic predictions, it is important to understand the
relationship between what is measured observationally (using
techniques such as that of eqn. A3) with what is predicted (which is
closer to the methods of eqns. 3 \& 4).  We thus have compared the
results of applying a velocity dispersion estimator based on spherical
averages (implicitly assuming a spherical isothermal model) to the
line-of-sight velocity dispersion $\sigma_{gal}$ obtained by applying
the point estimator given in appendix A (eq. A3), applying each to the
real-space FoF catalog of systems (our standard comparison sample).  
\vbox{%
\begin{center}
\leavevmode
\hbox{%
\epsfxsize=8.9cm
\epsffile{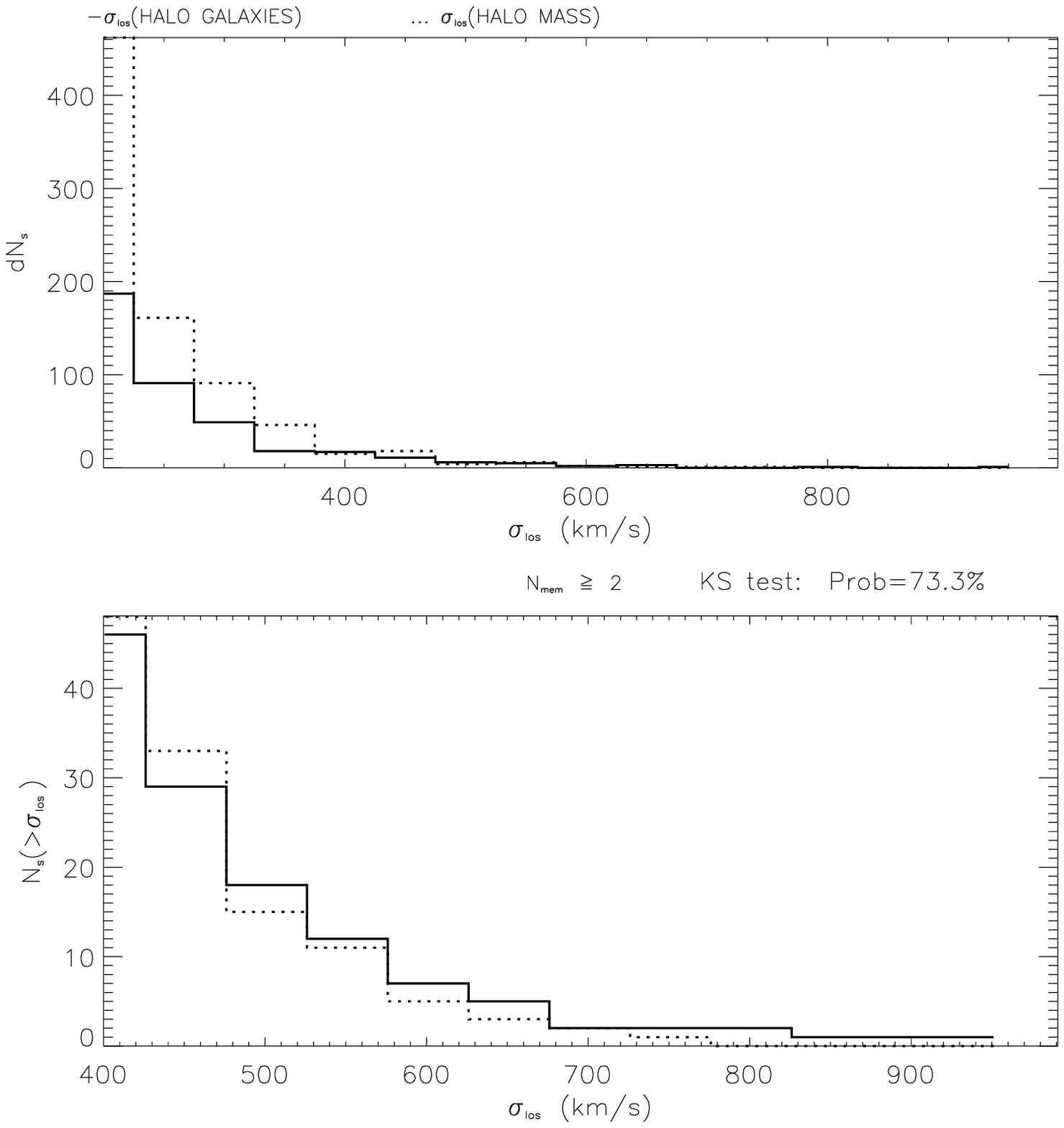}}
\begin{small}
\figcaption{%
Differential ({\em upper}) and cumulative ({\em lower}) velocity
dispersion distributions of systems identified via FOF in real space
for mock catalog \#1 and using two different methods of calculation:
applying the estimator described in appendix A (eq. A3) to all the
galaxies that form in the halo (solid line) and using a spherical
isothermal model to describe the halo mass distribution (dotted line).
Data are binned in velocity dispersion intervals of width 50 \kms.
Due to the offset between the two indicators (see Fig. \ref{fig22}),
we have re-mapped $\sigma_{los}$ inferred using galaxies by shifting each
estimated value lower by 50 \kms. A Kolmogorov-Smirnov test confirms
the consistency.
\label{fig23}}
\end{small}
\end{center}}

In figure \ref{fig22} we show the
scatter diagram between the two estimators applied to halos that
harbor at least 2 galaxies (regardless of their luminosity).  Compared
to figure \ref{fig21}, the scatter in figure \ref{fig22} is
substantial, since the groups found by FOF can be quite irregularly
shaped.  In addition to that scatter, systematic deviations from equality are apparent; in the low velocity regime ($\sigma_{gal} \leq 300$ \kms) we
have $\sigma_{halo}>\sigma_{gal}$ systematically, while for more massive
systems the spherical estimator is biased low with respect to
$\sigma_{gal}$.  Although the galaxy-based velocity estimator may be  less reliable  in the small-system
limit and the reality of that deviation less certain, there is  
a known tendency for  larger groups to be more irregularly shaped,
since FOF generally links galaxies  well outside $r_{200}$.
This in turn causes the two operational 
definitions of velocity dispersion to be biased in the same sense \citet{w01} found,
using a different spherical model to describe the  halo mass distribution.

These halo substructures are either real features or a problem caused
by the spurious tendency of the FOF algorithm to merge close,
pre-merging units. We can better understand the nature of this offset
and the influence of substructure on the internal cluster dynamics by
analyzing figure \ref{fig23}.  Here we plot the number distribution of
halos as a function of their differently estimated velocity
dispersions.  After shifting each $\sigma_{gal}$ estimate lower by 50
\kms, we note that the shape of the two distributions is approximately
the same, as is confirmed by the relatively high value of the
Kolmogorov-Smirnov probability parameter.  This simple re-mapping
strategy guarantees that at least in principle we can easily correct
for the velocity definition being used in modeling the distribution of
virialized systems as a function of their velocity dispersions.

\section{Conclusions}

The  Voronoi diagram  and its  dual, the  Delaunay  triangulation, are
among the  most useful data structures in  computational geometry. The
emphasis   of  this  paper   is  to   show  how   a  three-dimensional
implementation of these fundamental  graphs, which provide an explicit
representation of the relationships between neighboring points, can help
in  both identifying and  reconstructing a  complete sample  of galaxy
systems  in  a deep  flux-limited  redshift  survey,  with the  sample
selection  function   directly  defined  in  terms   of  the  velocity
dispersion parameter.
Our VDM algorithm uses locally specified clustering parameters, 
physically  determined on a cluster by cluster basis, without requiring  
arbitrarily  chosen  global linking parameters. Since it is based on a
neighboring relationship already encoded in the Delaunay complex, it is quite fast; 
only 5 minutes on  a modern  workstation  processes $\sim  15000$  
galaxies into  a catalog of groups and clusters.

We have used a family of artificial catalogs simulating the spatial,
velocity, and flux distributions of DEEP2 galaxies to study the
performance of our method, to test its stability under variation of
simple and physically justified clustering criteria, and to assess the
level of completeness of the resulting catalog of systems.

From applying the method to six independent mock catalogs  we conclude that:

i) The algorithm does not suffer from major distance-dependent effects
   and can be applied without any redshift-dependent correction to the
   parameters  governing  cluster  reconstruction over  the  wide
   redshift baseline  z=0.7-1.2 covered by the  DEEP2 redshift survey.
   The resulting  cluster catalog is fairly stable  under variations as
   great as 20\% of the selection parameters, and its members are identified
   in an  unbiased way for groups with 
   $\sigma_{los} \gtrsim  200$ \kms ~in a $I_{AB} \leq 23.5$ magnitude-limited survey.

ii) The recovered number density of systems as a bivariate function of
    velocity  dispersion and redshift  reproduces major features of the underlying real-space  distribution.  This function,  essentially free  from major
    statistical or  systematic uncertainties, can be  reliably used to
    constrain cosmological quantities as in \citet{new01}.

iii) The virial masses derived using samples of clusters complete in
 velocity dispersion match the actual cluster distribution poorly.
 Their values are biased high by the presence of interlopers which
 artificially increase the system radius.

iv) The necessity of avoiding overlapping spectra in the DEEP2 survey
    limits the sampling efficiency of spectroscopic target selection,
    especially in regions of the sky with higher than average surface
    density of galaxies.  Taking into account this target selection
    bias, we expect to identify and reconstruct a homogeneous sample
    of $\sim 270$ clusters per square degree with $N\sbr{mem}\geq 5$
    elements and velocity dispersion $\sigma\sbr{los} \gtrsim 200$
    \kms ~for a $\Lambda CDM$ cosmology.  Relaxing the richness
    condition, i.e.  only requiring $N\sbr{mem}\geq 2$, the unbiased sample would be a factor of 2 smaller, extending to
    $\sigma_{los} \sim 400$ \kms.

v) Finally, comparing the distribution properties of VDM reconstructed galaxy
systems and similar statistics of matter halos within a $\Lambda$CDM 
framework, we conclude that the recovered cluster sample is complete 
down to groups with $\sigma_{los} \gtrsim  400$ \kms. This limit sets the 
lower threshold for meaningful comparisons with theoretical models 
of structure formation.

The  DEEP2 survey  is  designed to  be  a comprehensive  study of  the
Universe  at  $z=1$.   Having  a  sample  of  high  redshift  galaxies
comparable  in quality  to those  available locally  will  enable many
independent tests of  the evolution of structure in  the Universe.  In
this study we demonstrate that we  expect DEEP2 to provide us not only
with a complete database of  galaxy positions and spectra, but also  with a
large, robust and complete  catalog of deep, optically selected galaxy
systems.

\acknowledgements  We  would like  to  thank  Guinevere Kauffmann  for
 allowing  us to use her semi-analytic  GIF simulations.  
A special thanks also to 
Martin  White and Chris Kochanek for enlightening discussions. 
We also    acknowledge   useful    conversations    with   S.    Borgani,
J. di Francesco, and E.  Scannapieco.
This work  is supported by  National  Science  
Fundation Grant No. AST-0071048 and by equipment donated by Sun Microsystems.

\appendix

\section{Definitions of system properties}

The expression relating cosmological ($z$)  and observed ($z_o$)
redshifts is, in 
first order approximation, \citep{har74}
\begin{equation}
z = z_0+\frac{v_p}{c} (1+z_0)
\end{equation}
\noindent where $v_p$ (km s$^{-1}$) is the peculiar velocity along the line of sight.
In the analysis of the $\Lambda$CDM mock catalogs, the relations between
$z_0$ and $r$ is determined using  the comoving distance formula
\begin{equation}
r = \frac{c}{H_0}\int_{0}^{z}\frac{dz}
{[\Omega_{0,m}(1+z)^{3}+\Omega_{\Lambda}-\Omega_k (1+z)^2]^{1/2}}. 
\end{equation}

\noindent The {\em line of sight velocity dispersion} in
a group with $N_{mem}$ members each at redshift $z_i$, corrected for relativistic effects \citep{har74}, is
\begin{equation}
\sigma_{\rm los} = \frac{1}{1 + <z>} \sqrt{ 
\frac{\sum_{i=1}^{N_{\rm mem}} (cz_i - <cz>)^2}{(N_{\rm mem} - 1)} } .
\end{equation}

A formal estimate of the standard error in
$\sigma_{\rm los}$ is then given by \citep{bev69}:
\begin{equation}
\sigma(\sigma_{\rm los}) = \frac{1}{2} \sigma_{\rm los} \sqrt{ \frac{2}{N_{\rm mem}} 
\left( 1-\frac{1}{N_{\rm mem}} \right) }.
\end{equation}

\noindent The {\em mean pairwise separation} in a group at distance $z_{gr}$ 
may be calculated according to the approximate formula
\begin{equation}
R_{\rm p} \equiv <|{\bf r}_{ij}|>  \approx
 \frac{4}{\pi} d_{\rm ang}(z_{gr}) <\theta_{ij}> ,
\end{equation}
where 
\begin{equation}
<\theta_{ij}> \equiv \frac{\sum_{i} \sum_{j>i} \theta_{ij}}{N_{\rm pair}} , 
\end{equation}
$N_{\rm pair}$ is the number of distinct galaxy pairs in the
group ($N_{mem}(N_{mem}-1)/2$), 
$d_{\rm ang}(z_{gr})$ is the comoving angular distance of the group
at the given redshift,
and 
$\theta_{ij}$ is the angular separation between group members $i$ and
$j$. \citet{tuc00} estimate the rms error in $R_{\rm p}$ to be
\begin{equation}
\sigma_{R_{\rm p}} = \frac{R_{\rm p}}{<\theta_{ij}>}
\sqrt{
\frac{ N_{\rm pair} \sum_{i} \sum_{j>i} \left(  \theta_{ij} \right)^2 - 
       \left( \sum_{i} \sum_{j>i} \theta_{ij}  \right)^2 }{
       \left( N_{\rm pair} - 1 \right)N_{\rm pair}^2 }}.
\end{equation}

\noindent The group {\em harmonic radius} is approximately given  by the
formula
\begin{equation}
R_{\rm h} \equiv <|{\bf r}_{ij}|^{-1}>^{-1} \approx \frac{\pi}{2}  \frac{d_{ang}(z_{gr})}
{<\theta_{ij}^{-1}>}, 
\end{equation}
while its error will be 
\begin{equation}
\sigma_{R_{\rm h}} =\frac{R_{\rm h}}{<\theta_{ij}^{-1}>}
                      \sqrt{
\frac{ N_{\rm pair} \sum_{i} \sum_{j>i} \left( \theta_{ij}^{-1} \right)^2 - 
       \left( \sum_{i} \sum_{j>i} \theta_{ij}^{-1}  \right)^2 }{
       \left( N_{\rm pair} - 1 \right) N_{\rm pair}^2 }. 
} 
\end{equation}

\noindent The {\em crossing time} for a group, in Hubble time units,
is defined as in \citet{nw87}:
\begin{equation}
\frac{t_{\rm cr}}{t\sbr{H}} = \frac{2 R_{\rm h}}{\sqrt{3} \sigma_{\rm los}} \left[ 1 \pm
\sqrt{ \frac{\sigma_{R_{\rm h}}^2}{R_{\rm h}^2} + 
       \frac{\sigma_{\sigma_{\rm los}}^2}{\sigma_{\rm los}^2}} \; \right].
\end{equation}

Comparing this time to the time needed for a homogeneous sphere
to collapse out of the Hubble flow, we can formally conclude that a system 
is virialized if $t_{cr} <0.34\;t\sbr{H}$. 

\noindent The group's virial mass is estimated to be  
\begin{equation}
M_{\rm vir} = \frac{6 \sigma_{\rm los}^2 R_{\rm h}}{G} \left[ 1 \pm 
\sqrt{ \frac{4 \sigma_{\sigma_{\rm los}}^2}{\sigma_{\rm los}^2} + 
       \frac{\sigma_{R_{\rm h}}^2}{R_{\rm h}^2} } \; \right],
\end{equation}
where $G$ is the gravitational constant.

A more stable mass estimator, less sensitive to interlopers, is the median
mass defined as \citep{htb85}.

\begin{equation}
M\sbr{med} = \frac{f}{G} d_{ang}(z_{gr}) Med_{ij} [(v\sbr{i}-v\sbr{j})^2 
           \theta\sbr{ij}],
\end{equation}

\noindent where $v$ is the  velocity of the cluster members with respect to
the cluster centroid, corrected for relativistic effects, and 
where $f$ is a dimensionless fudge factor determined from  numerical 
experiments and whose value is set to 6.5.

%Following \citet{nw87}
%we define the density contrast of a group, i.e. 
%its number density 
%enhancement over the local mean, in terms of the system mean pairwise 
%separation:
%
%\begin{equation}
%\frac{\delta \rho}{\rho}=\frac{3N_{mem}}{4 \pi R_{\rm p}^3 \int_{L_{lim(d)}}^{\infty}
%\Phi(L)dL}-1,
%\end{equation}
%where $\Phi(L)$ is the luminosity function of our mock simulation.

We  reconstruct the total luminosity of each system using the 
weighting scheme of \citet{mhg01}, i.e. we write  the total
expected luminosity of the system as
\begin{equation} L\sbr{s}(r) = L\sbr{obs}(r)+ \tilde{L}(r) = L\sbr{obs}(r)
w_{L}(r) \label{mco}, \end{equation} 
where $\tilde{L}$ denotes the unseen luminosity from galaxies below
the magnitude limit and $w_{L}(r)$ is the {\em luminosity-density
weighting function} 
\begin{equation} w_{L}=\frac{\int_{0}^{\infty} L \phi(L) dL}
{\int_{L\sbr{lim}(d)}^{\infty} L \phi(L)dL}.  \label{lsf} \end{equation}
A formal estimate of the rms error is obtaining by summing in quadrature 
the errors of the single luminosities $L_i$ of the group members as follows:
\begin{equation} 
\sigma_{L_s}=w_{L} N_{mem}\sqrt{\frac{<L_i^2>-<L_i>^2}{N_{mem}-1}}.
\end{equation}

%A different way of correcting for the missing light has also been 
%implemented:
%\begin{equation}
%\tilde{L} = N\sbr{mem} \frac{\int_{0}^{L\sbr{lim}(d)} L\phi(L)dL}
%{\int_{L\sbr{lim}(d)}^{\infty} \phi(L)dL.}
%\end{equation}

% REFERENCES
\bibliographystyle{apj}
\bibliography{paper, book}

% TABLES

\scriptsize{
\begin{deluxetable}{clcccccc}
\tablewidth{0pc}
\tablecaption{Median group properties in real and redshift space mock catalogs\label{tab1}}
\tablehead{
\colhead{Catalog} &
\colhead{ }   &
\multicolumn{2}{c}{Sample I\tablenotemark{a}} &
\multicolumn{2}{c}{Sample II\tablenotemark{b}} &
\multicolumn{2}{c}{Sample III\tablenotemark{c}} \\
\colhead{} &
\colhead{} &
\colhead{r-space} &
\colhead{z-space} &
\colhead{r-space} &
\colhead{z-space} &
\colhead{r-space} &
\colhead{z-space} \\
}
\startdata
&$N\sbr{s}$                              &1678 &1517    & 571  &528 &251 & 243  \\
&Med$(N\sbr{mem})$                       &2    & 4      & 4    &6   &7   & 7     \\
&Med$(R\sbr{P})$                         &0.41 &0.85    & 0.53 &0.9 &0.66&1.07   \\
&Med$(R\sbr{H})$                         &0.42 &0.71    & 0.48 &0.68&0.54&0.80    \\
&Med$(t\sbr{cr})$ (Hubble time)          &0.33 &0.55    & 0.17 &0.24&0.18&0.25    \\
Mock \#1&Med$(\sigma\sbr{los})$                  &136  &133     & 301  &302 &328 & 327 \\
&Med$(M\sbr{v}) (10^{13} M_{\odot}) $    &0.9  &1.64    & 5.9  &8.3 &7.7 & 11 \\
&Med$(M\sbr{v}) (10^{13} M_{\odot}) $    &1.2  &1.9     & 6.9  &8.8 &7.4	& 11 \\
&Med$(L\sbr{v}) (10^{11} L_{\odot}) $    &0.7  &1.1     & 1.3  &1.9 &2.8	& 2.2 \\
&Med$(M/L) (M_{\odot}/L_{\odot})$        &119  &119     & 378  &362 &264	& 450  \\
\hline

&$N\sbr{s}$                              &1822 &1764    & 624  &578 &287   &290  \\
&Med$(N\sbr{mem})$                       &2    & 4      & 4    &6   &8     & 7 \\
&Med$(R\sbr{P})$                         &0.38 &0.85    & 0.49 &0.87&0.64  &1.05 \\
&Med$(R\sbr{H})$                         &0.38 &0.69    & 0.44 &0.66&0.49  &0.81 \\
&Med$(t\sbr{cr})$ (Hubble time)          &0.31 &0.53    & 0.16 &0.23&0.16  &0.27 \\
Mock \#3&Med$(\sigma\sbr{los})$          &135  &130     & 297  &305 & 335  &317 \\
&Med$(M\sbr{v}) (10^{13} M_{\odot}) $    &0.9  &1.4     & 5.4  &8.4 & 7.32 &11.4 \\
&Med$(M\sbr{v}) (10^{13} M_{\odot}) $    &1.1  &1.6     & 6.6  &9.5 & 7.37 &10.6 \\
&Med$(L\sbr{v}) (10^{11} L_{\odot}) $    &0.7  &1.1     & 1.3  &2.2 & 2.90 &2.34 \\
&Med$(M/L) (M_{\odot}/L_{\odot})$        &106  &112     & 337  &350 & 241  &457 \\
\enddata
\tablenotetext{a}{All groups; i.e. systems with $N\sbr{mem} \geq 2$ members}
\tablenotetext{b}{Only groups with $\sigma_{los} \geq 200$ \kms}
\tablenotetext{c}{Only groups with $\sigma_{los} \geq 200$ \kms ~and 
$N\sbr{mem} \geq 5$ extracted from the mock catalogs after applying the DEEP2 target 
selection criteria}

\end{deluxetable}

\end{document}